\documentclass[letter]{aa}
\usepackage[varg]{txfonts}
\begin{document}

\title{Magnetically aligned straight depolarization canals \\ and the rolling Hough transform}
\titlerunning{LOFAR depolarization canals and RHT}
\author{Vibor Jeli\'c\inst{1}, David Prelogovi\'c\inst{2}, Marijke Haverkorn\inst{3}, Jur Remeijn\inst{3}, and Dora Klind\v{z}i\'c\inst{2}}
\date{\today}
\institute{Ru{\dj}er Bo\v{s}kovi\'{c} Institute, Bijeni\v{c}ka cesta 54, 10000 Zagreb, Croatia,
\email{vibor@irb.hr}
  \and Department of Physics, Faculty of Science, University of Zagreb, Bijeni\v{c}ka cesta 32, 10000 Zagreb, Croatia
  \and Department of Astrophysics / IMAPP, Radboud University Nijmegen, PO Box 9010, 6500 GL Nijmegen, the Netherlands}
\abstract
{}
{We aim to characterize the properties of the straight depolarization canals detected in the Low Frequency Array (LOFAR) polarimetric observations of a field centered on the extragalactic source  \object{3C 196}. We also compare the canal orientations with magnetically aligned H{\sc i} filaments and the magnetic field probed by polarized dust emission.}
{We used the rolling Hough transform (RHT) to identify and characterize the orientation of the straight  depolarization  canals in radio polarimetric data and the filaments in H{\sc i} data.}
{The majority of the straight depolarization canals and the H{\sc i} filaments are inclined by $\sim10^\circ$ with respect to the Galactic plane and are aligned with the plane-of-sky magnetic field orientation probed by the \emph{Planck} dust polarization data. The  other  distinct  orientation,  of $-65^\circ$ with respect to the Galactic plane, is associated with the orientation of a bar-like structure  observed  in  the 3C 196 field  at  350 MHz.}
{An alignment  between  three  distinct  tracers  of  the  (local) interstellar medium (ISM) suggests that  an ordered  magnetic field plays a crucial role in confining different ISM phases. The majority of the straight depolarization canals are a result of a projection of the complicated 3D distribution of the ISM. The RHT analysis is a robust method for identifying and characterizing  the  straight  depolarization  canals  observed in radio-polarimetric data.}

\keywords{ISM:general, magnetic fields, structure - radio continuum: ISM - techniques: interferometric, polarimetric} 

\maketitle

\section{Introduction}\label{sec:intro}
Observations at low radio frequencies ($\sim100-200~{\rm MHz}$) show a very rich morphology of Galactic synchrotron emission in polarization \citep{jelic14, jelic15, lenc16, vaneck17}. The discovered structures are unraveled by the rotation measure (RM) synthesis \citep{brentjens05}, a  technique in radio polarimetry that separates the observed polarized emission according to the degree of Faraday rotation it has experienced. This allows us then to study and map the relative distribution of the intervening magneto-ionic interstellar medium (ISM) as a function of Faraday depth, that is, to perform a so-called Faraday tomography. The Faraday depth ($\Phi$) is defined as
\begin{equation}
\frac{\Phi}{[{\rm rad~m^{-2}}]}=0.81\int\frac{n_e}{[{\rm cm^{-3}}]}\frac{B_{||}}{[{\rm \mu G}]}\frac{{\rm d}l}{[{\rm pc}]},
\end{equation}
where $n_e$ is electron density and $B_{||}$ is the magnetic field component parallel to the line of sight,  ${\rm d}l$.  The integral is taken over the entire path from the source to the observer. A positive Faraday depth denotes a magnetic field component pointing toward the observer. A negative Faraday depth denotes a magnetic field component pointing away from the observer.

A Faraday tomography of the Milky Way was first achieved using Westerbork Synthesis Radio Telescope (WSRT) observations at 350 MHz \citep[e.g.,][]{brentjens07} and later at 150 MHz \citep{bernardi09, bernardi10, iacobelli13a}. However, its true power came to the fore when new low-frequency interferometers came online: the Low Frequency Array \citep[LOFAR,][]{haarlem13} and the Murchison Widefield Array \citep[MWA,][]{tingay13}.  Their wide frequency coverage, accompanied with a high spectral resolution, resulted in an exquisite sensitivity and resolution in Faraday depth ($\sim 1~{\rm rad~m^{-2}}$), which is an order of magnitude higher than at 350 MHz. Hence, Faraday tomography at low radio frequencies is sensitive to low column densities of magnetized plasma (e.g., mostly warm ionized medium, WIM) that in most cases cannot be detected at higher radio frequencies. This is supported by the observations at low radio frequencies described above.

Based on LOFAR polarimetric observations, the Faraday tomography of a field centered at \object{3C~196} shows an astonishing variety of features \citep{jelic15}. The most striking structure is a straight filament of likely ionized gas that is a few degrees long and is located somewhere within the Local Bubble \citep{lallement14}, which displaces the background synchrotron emission in Faraday depth. Together with another filamentary structure observed in the same field, it correlates with the cold filamentary H{\sc i} structures \citep{kalberla16b} and the magnetic field orientation probed by the \emph{Planck} observations of the dust emission in polarization \citep{zaroubi15}. \citet{kalberla17} also found a similar alignment between the different ISM components in two other fields in the sky, Horologium and Auriga,  which were observed with the WSRT at 350 MHz \citep{haverkorn03a,haverkorn03b}.

The boundaries of the features observed in the 3C~196 field are clearly defined by depolarization canals. They are apparent in an image showing the highest peak of the Faraday depth spectrum in polarized intensity at each spatial pixel \citep[see Fig. 6 in][]{jelic15}. They resemble the characteristics of depolarization canals described by \citet{haverkorn04}. Most of them are associated with beam depolarization that is due to discontinuity in the polarization angle between two regions. Others might be caused by differential Faraday rotation, which completely depolarizes emission along the line of sight. A real puzzle, however, is that most of them are extraordinary straight. As discussed in \citet{jelic15}, their straightness might be the result of a fortunate projection of the complicated three-dimensional morphology of the magnetic field and the ISM (e.g., the folding of sheets, and loops and shocks), or it might be associated with interactions of close-by moving stars and the ISM. 

We here explore some of these possibilities further by characterizing the orientation and position of the straight depolarization canals observed in the 3C 196 field using the rolling Hough transform \citep[RHT, ][]{clark14}. We then compare their orientations with H{\sc i} structures detected in the same field \citep{kalberla16b} and with the magnetic field probed by the \emph{Planck} satellite \citep[][]{planckXIXinter}. Observational data used for this work are detailed in Section~\ref{sec:obs},  the method of the RHT analysis is described in Section~\ref{sec:RHT}, and its results are presented in Section~\ref{sec:res}. The paper concludes with a discussion and summary in Section~\ref{sec:dis}.

\section{Observational data}\label{sec:obs}
We used a publicly available Faraday depth cube of 3C 196 field, given in polarized intensity \citep{jelic15}\footnote{http://cdsweb.u-strasbg.fr/cgi-bin/qcat?J/A+A/583/A137}. The cube was synthesized based on the LOFAR High Band Antennas observation L80508, described in \citet{jelic15} and covers the frequency range from 115 MHz to 189 MHz. The Faraday depth cube has an angular resolution of $\sim3.75~{\rm arcmin}$. The resolution in Faraday depth is $\sim1~{\rm rad~m^{-2}}$, and the noise is $70~{\rm \mu Jy~PSF^{-1} ~RMSF^{-1}}$.

\begin{figure}[!t]
    \includegraphics[width=\linewidth]{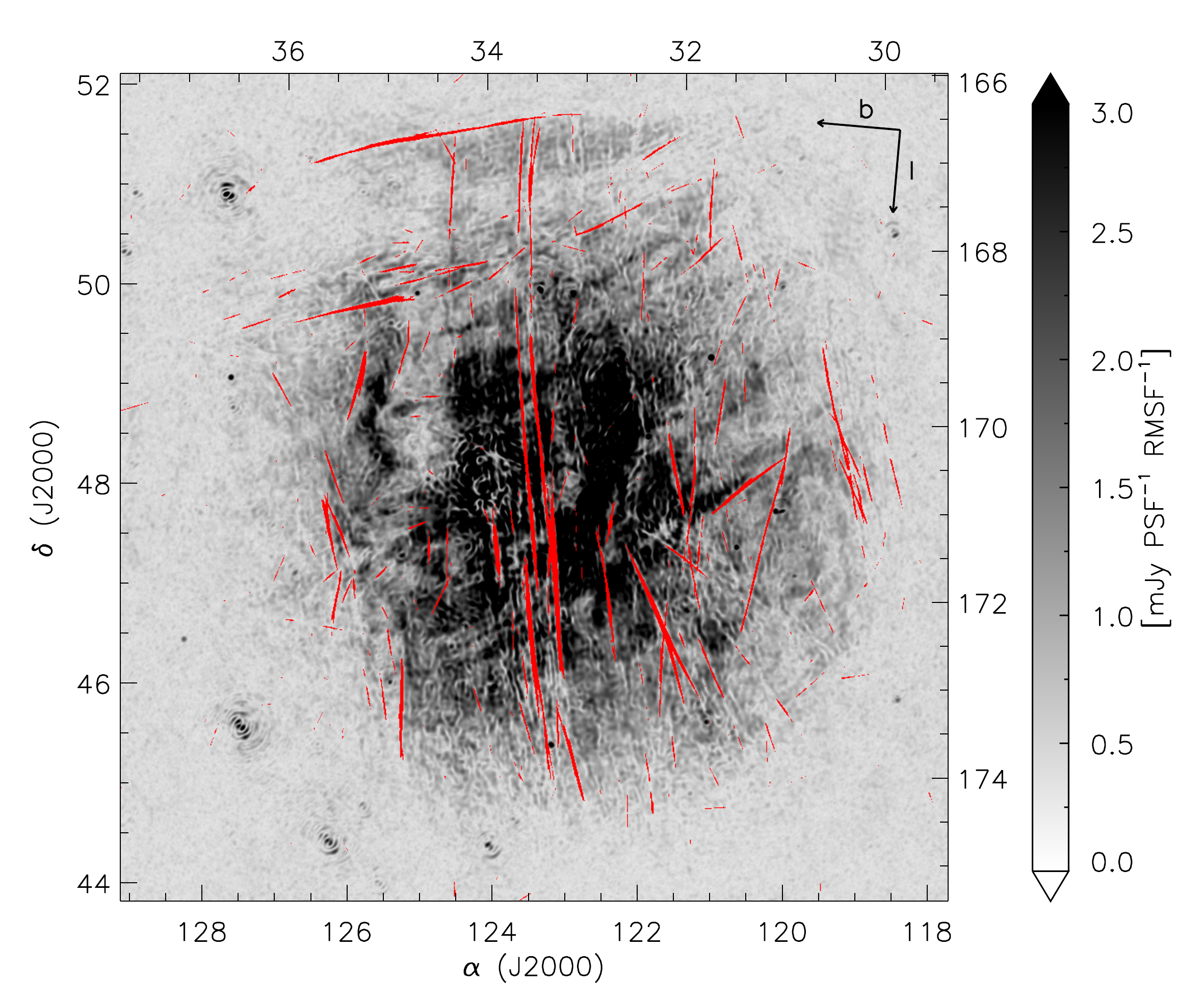}
    \caption{Example of the RHT backprojection (red lines) visualized on an inverted image of the highest peak of the Faraday spectrum in polarized intensity at each spatial pixel for the 3C 196 field. The input parameters for the RHT are $D_K = 4'$, $D_W = 50'$, and $Z = 0.8$.}
    \label{fig:RHTimage}
\end{figure}

To identify straight depolarization canals, we have recreated an image showing the highest peak of the Faraday depth spectrum in polarized intensity at each spatial pixel \citep[see fig. 6 in][ and Fig.~\ref{fig:RHTimage}]{jelic15}. To this image, we then applied the RHT described in Section~\ref{sec:RHT}.

For the H{\sc i} distribution in 3C 196 field, we used publicly available data of the Galactic Effelsberg–Bonn H{\sc i} Survey \citep[EBHIS, ][]{winkel16}\footnote{http://cdsarc.u-strasbg.fr/viz-bin/qcat?J/A+A/585/A41}. The EBHIS data has an angular resolution of $10.8~{\rm arcmin}$ and a spectral resolution of $1.44~{\rm km~s^{-1}}$.

In our analysis we also used the plane-of-sky magnetic field orientation traced by a polarization angle of dust emission rotated by $90^\circ$. This we inferred directly from the \emph{Planck} 353 GHz polarization maps given in Stokes Q,U \citep{planckXIXinter,planck2015I}. The maps are publicly available at the Planck Legacy Archive\footnote{http://pla.esac.esa.int}. Because the signal-to-noise ratio of the observed polarized dust emission in the 3C 196 field is lower, we used the \emph{Planck} data smoothed to a resolution of $15~{\rm arcmin}$. 

\section{Rolling Hough transform}\label{sec:RHT}
The rolling Hough transform \citep[RHT,][]{clark14}\footnote{https://github.com/seclark/RHT} is a modification of the Hough transform algorithm \citep{hough96}, which is widely used to detect discrete straight lines. The RHT quantifies each pixel of an image with respect to its surroundings by encoding the probability that it is a part of a coherent linear structure. 
 In astronomy it was initially used to characterize magnetically aligned filaments in H{\sc i} data \citep{clark14}. We here use it to characterize straight depolarization canals observed in the radio polarimetric data.

The output of the RHT is the function $R(\theta,x,y)$, where $\theta$ is an angle of a parameterized straight line and $x,y$ are coordinates of a pixel in an image. To visualize this result, the backprojection is obtained by integrating $R(\theta,x,y)$ over $\theta$. To highlight specific linear features in the RHT output, three input parameters can be used: (i) the smoothing kernel diameter ($D_K$), which controls a suppression of large-scale structures in an image; (ii) the window diameter ($D_W$), which defines a minimum length of linear structures; and (iii) the probability threshold ($Z$), which defines a lower cut of the stored outputs $R(\theta,x,y)$. 

\begin{figure*}[!th]
    \centering
    \begin{tabular}{c|cccc}
        & $D_W$ & $30'$  & $40'$ & $50'$ \\\hline
        $D_K$ &  &  &  & \\
        $4'$  &
        & \begin{minipage}{.2\textwidth}
              \includegraphics[width=\linewidth]{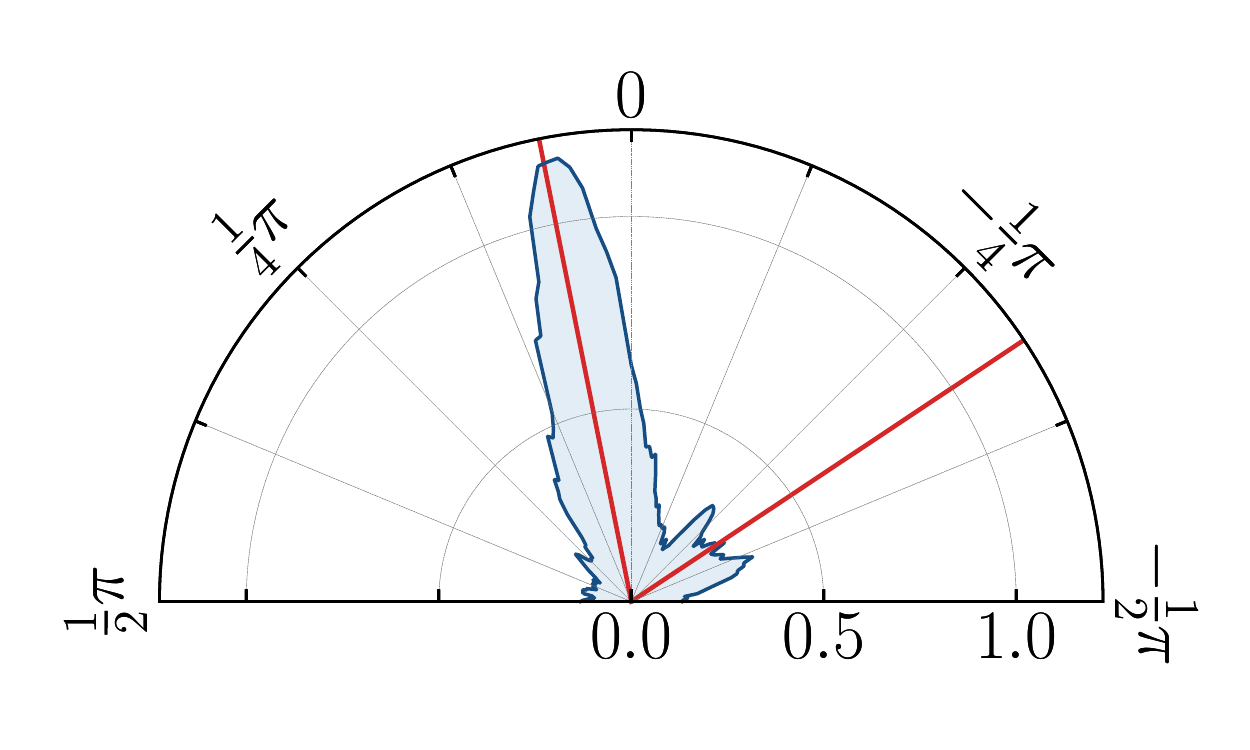}
              \end{minipage}
        & \begin{minipage}{.2\textwidth}
              \includegraphics[width=\linewidth]{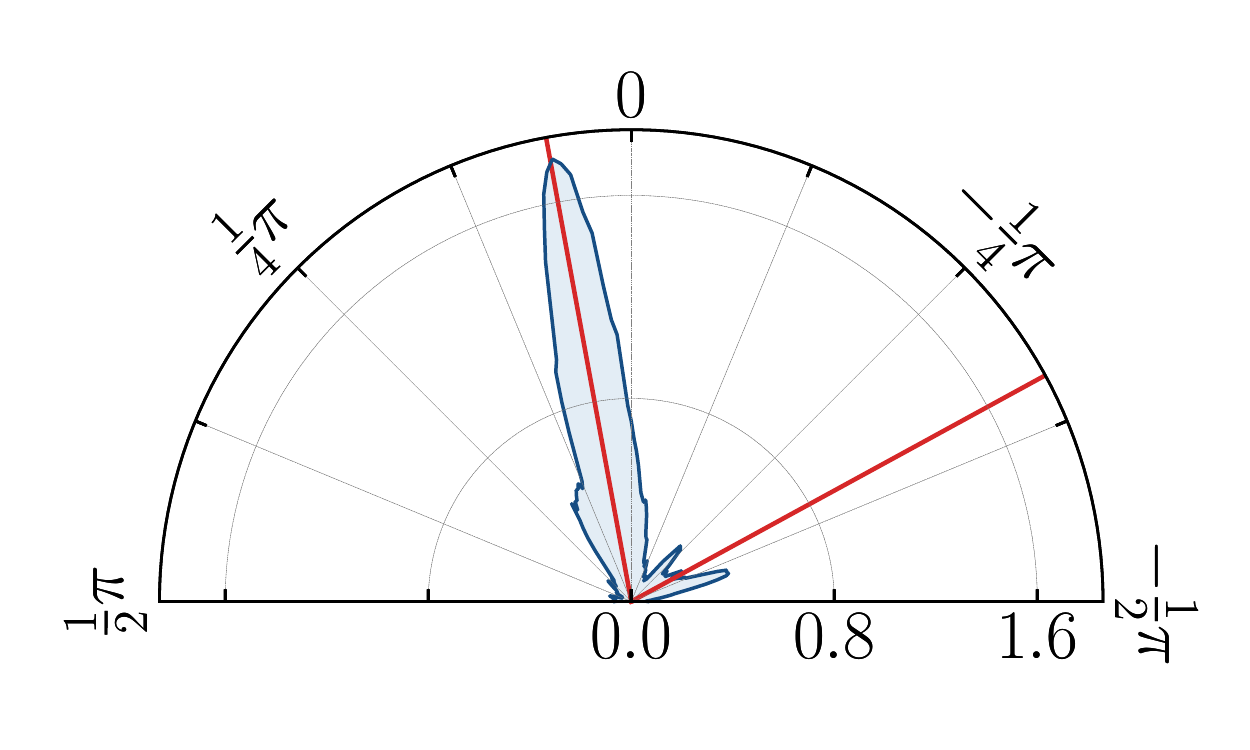}
              \end{minipage}
        & \begin{minipage}{.2\textwidth}
              \includegraphics[width=\linewidth]{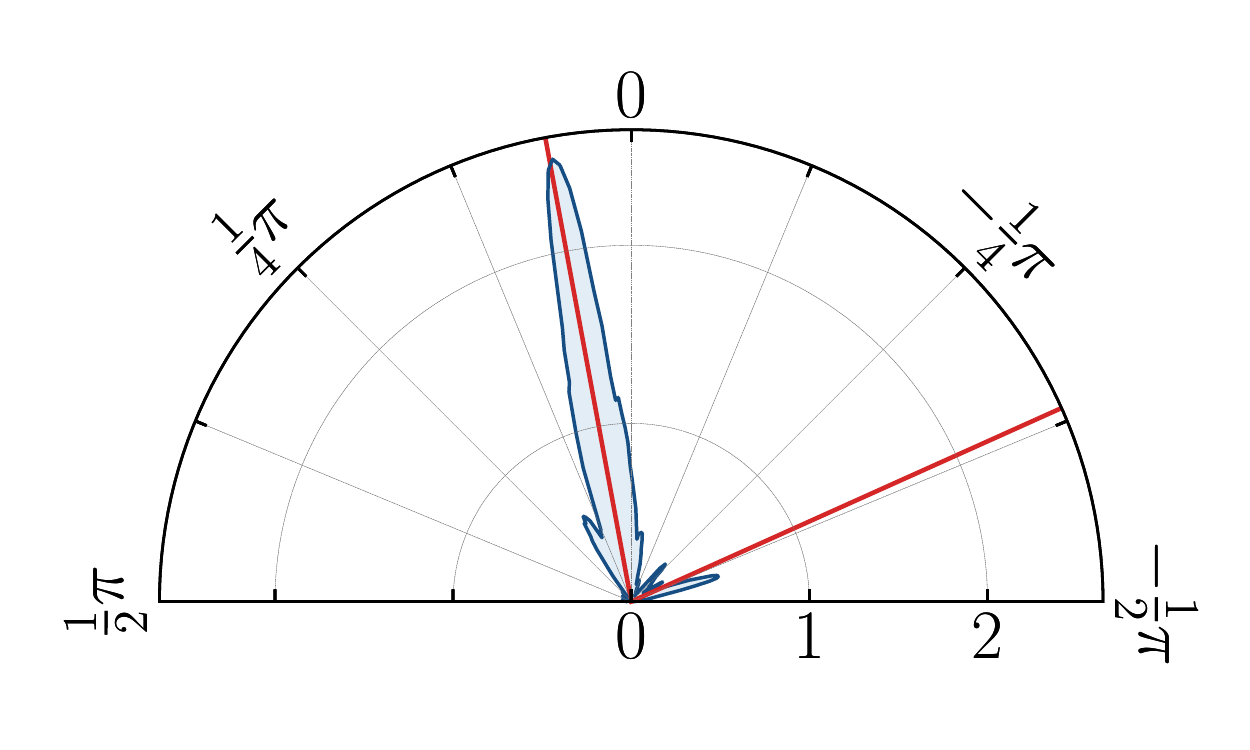}
              \end{minipage}
        \\
          $8'$ &
        & \begin{minipage}{.2\textwidth}
              \includegraphics[width=\linewidth]{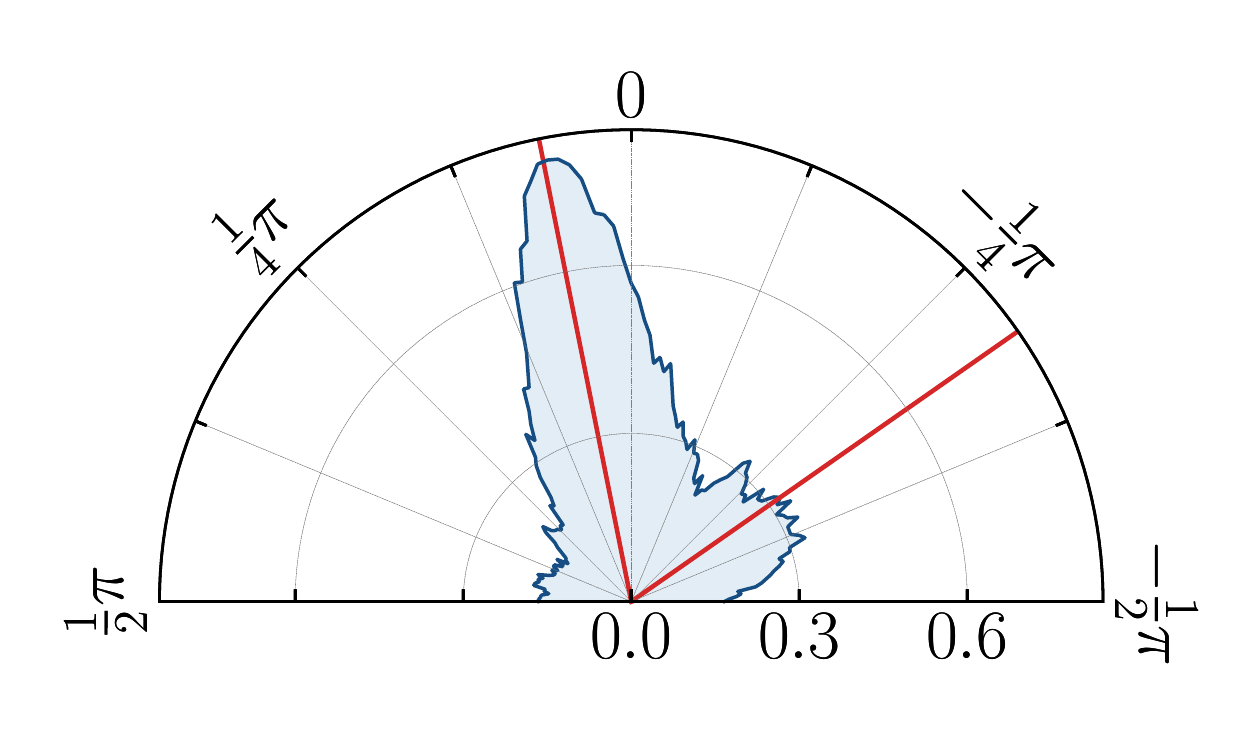}
              \end{minipage}
        & \begin{minipage}{.2\textwidth}
              \includegraphics[width=\linewidth]{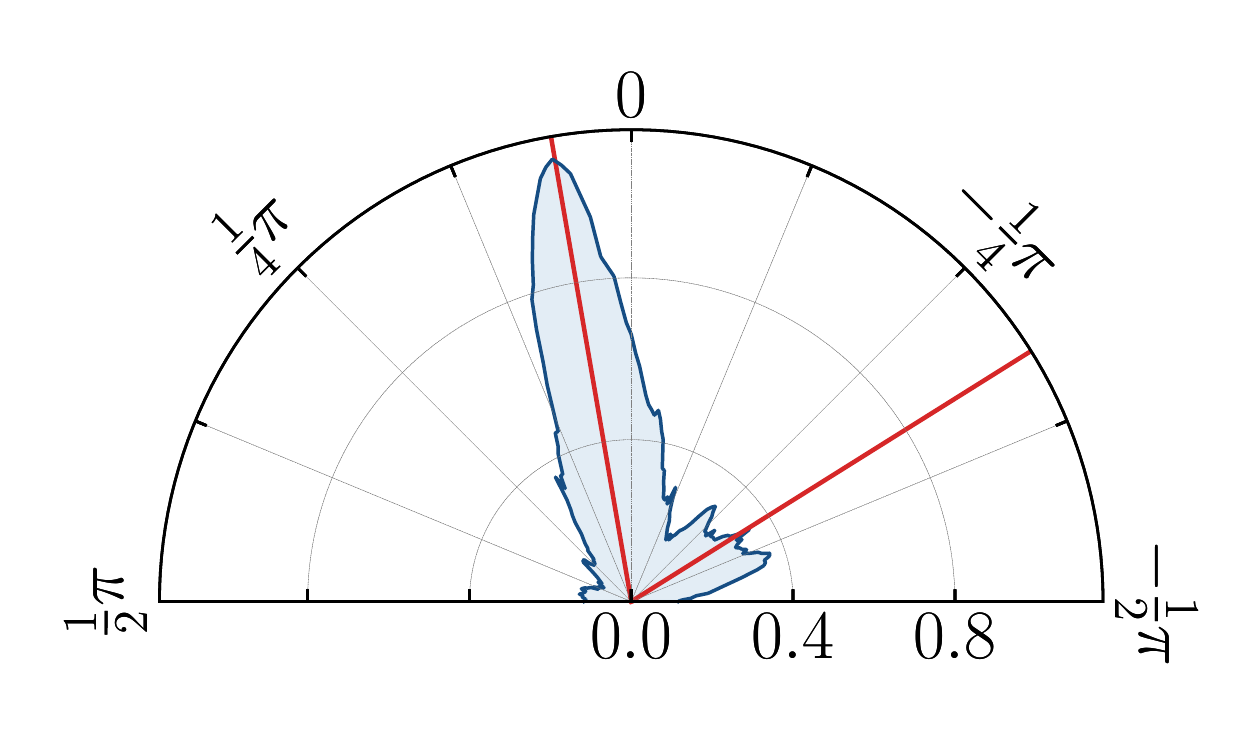}
              \end{minipage}
        & \begin{minipage}{.2\textwidth}
              \includegraphics[width=\linewidth]{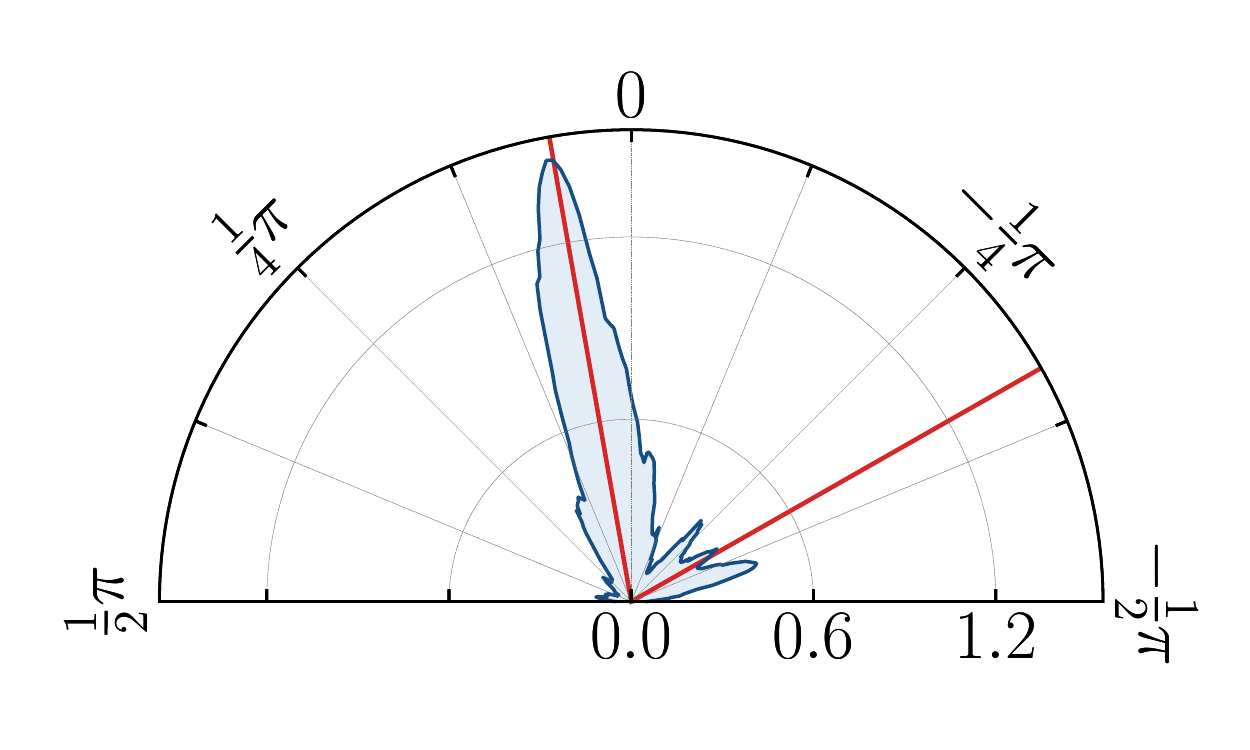}
              \end{minipage}
    \end{tabular}
    \caption{Relative orientation of the straight depolarization canals detected in the 3C 196 field, given for the RHT input parameter $Z=0.8$ and different values of $D_K$ and $D_W$. The plots are normalized such that $\int_{-\pi/2}^{\pi/2} \tilde{R}(\theta) \, d\theta = 1$. Red lines mark the means of two distinctive orientations in the overall distribution. The alignment with the Galactic plane is at $0$.}
    \label{fig:RHT_comp}
\end{figure*}

To quantify the relative orientation of linear structures  (identified with the RHT) with respect to the Galactic plane, we followed \citet{clark14} and defined a metric
\begin{equation}\label{eq:intensity_function}
    \tilde{R}(\theta) =\frac{1}{\mathcal{N}} \iint R(\theta,x,y)\,dx\,dy,
\end{equation}
where the integral goes over a region of an image, and $\mathcal{N}$ is a normalization factor that is commonly defined as $\int_{-\pi/2}^{\pi/2} \tilde{R}(\theta) \, d\theta = 1$. The results are then visualized on a half-polar plot, such that the alignment with the Galactic plane is at $0$ and the orthogonal alignment is at $\pm{\pi}/{2}$, following the counterclockwise direction.

When calculating a mean $\left<\theta\right>$ and a spread $\delta\theta$ of the distribution, we make a full circle projection of the half-polar plot and visualize every point as a vector with length $\tilde{R}^2 \, d\theta$. Integrating over the whole plane gives us a total vector $S$:
\begin{equation}
    S = \frac{\int_{-\pi/2}^{\pi/2} \tilde{R}^2(\theta) \,  e^{2i\theta} \, d\theta}
             {\int_{-\pi/2}^{\pi/2} \tilde{R}^2(\theta) \, d\theta} \, .
\end{equation}
The direction of $S$ and its value are the measures of $\left<\theta\right>$ and $\delta\theta$ in the following way:
\begin{align}
   \left< \theta \right> \,  &= \, \frac{1}{2} \mathrm{Arg} (S) \, = \, \frac{1}{2} \mathrm{arctan2} \left(\, \mathrm{Im}(S),\,  \mathrm{Re}(S)\, \right) \, ,\\
   \delta\theta \, &= \, \frac{1}{2} \sqrt{\ln\left(1/|S|^2\right)} \, ,
    \label{eq:theta_mean_deviation}
\end{align}
where the function $\mathrm{arctan2}$ returns the polar angle of a vector in Cartesian coordinates\footnote{ $\mathrm{arctan2}(y, x) = \arctan \left(\frac{y}{x}\right) + \pi \cdot \left( 1 - H(x) \right) \left( 2 H(y) - 1 \right) \, \in \left<-\pi, \pi\right] \, ,$ where $H(x)$ is the Heaviside step function ($H(0)=1$). $1/2$ in front is due to the full-circle projection.}.

\section{Results}\label{sec:res}
\subsection{Orientation of the straight depolarization canals}

To identify and characterize the straight depolarization canals observed in the 3C 196 field, we applied the RHT on an inverted image showing the highest peak of the Faraday spectrum in polarized intensity. An example of the RHT backprojection is shown in Fig.~\ref{fig:RHTimage} using the red lines. Almost all straight depolarization canals are well identified visually.

To conduct an exploration of the RHT input parameter space, we varied the smoothing kernel diameters $D_K$ from 2' to 8', the rolling window diameters $D_W$ from 20' to 80', and the intensity thresholds $Z$ from 0.5 to 0.8. The RHT backprojections were then visually compared, as were the half-polar plots quantifying the relative orientation of the depolarization canals with respect to the Galactic plane. 

The results of all parameter combinations visually agree with each other in terms of the main linear features in backprojection and their orientations. Smaller differences are mainly related to low-intensity background, when lower threshold values were used. Since this is also computationally very expensive, we fixed threshold to $Z=0.8$ for the remainder of the work.

Figure~\ref{fig:RHT_comp} shows a representative sample of the results for $Z=0.8$ and different values of $D_K$ and $D_W$. The similarities between the plots are evident. Two distinctive orientations are visible in the overall distribution. The first has a mean of $\left<\theta\right>=10^\circ$ and a spread of $\delta\theta=\pm6^\circ$ with respect to the Galactic plane, and the mean and spread of the second are $\left<\theta\right>=-65^\circ$ and $\delta\theta=\pm13^\circ$ respectively. These values are given for the RHT input parameters $Z=0.8$, $D_K=50'$ , and $D_W=8'$. 

The larger kernel diameters $D_K$ and smaller rolling window diameters $D_W$ result in a wider spread of the relative orientation of the depolarization canals. The reason behind this is rather simple. When a wider kernel is used, the large scales are suppressed and more pixels are collected in a bitmask image. In combination with a small rolling area, shorter lines start to contribute, acting as a background noise. This causes the overall distribution to become wider. However, the mean values $\left< \theta \right>$ of the two distinctive orientations remain roughly consistent across the input parameter space. This indicates that the longest depolarization canals dominate the distribution, and their orientations are similar.

\subsection{Orientation of the \emph{H{\sc i}} filaments}
The relative orientation of the H{\sc i} filaments in the 3C 196 field were obtained by applying the RHT on the H{\sc i} brightness temperature at each velocity channel of the EBHIS data. We then summed the results for the velocity range from $-11.5~{\rm km~s^{-1}}$ to $+18.1~{\rm km~s^{-1}}$, where the filaments are most evident. 

\begin{figure}[t]
\centering \includegraphics[width=.75\linewidth]{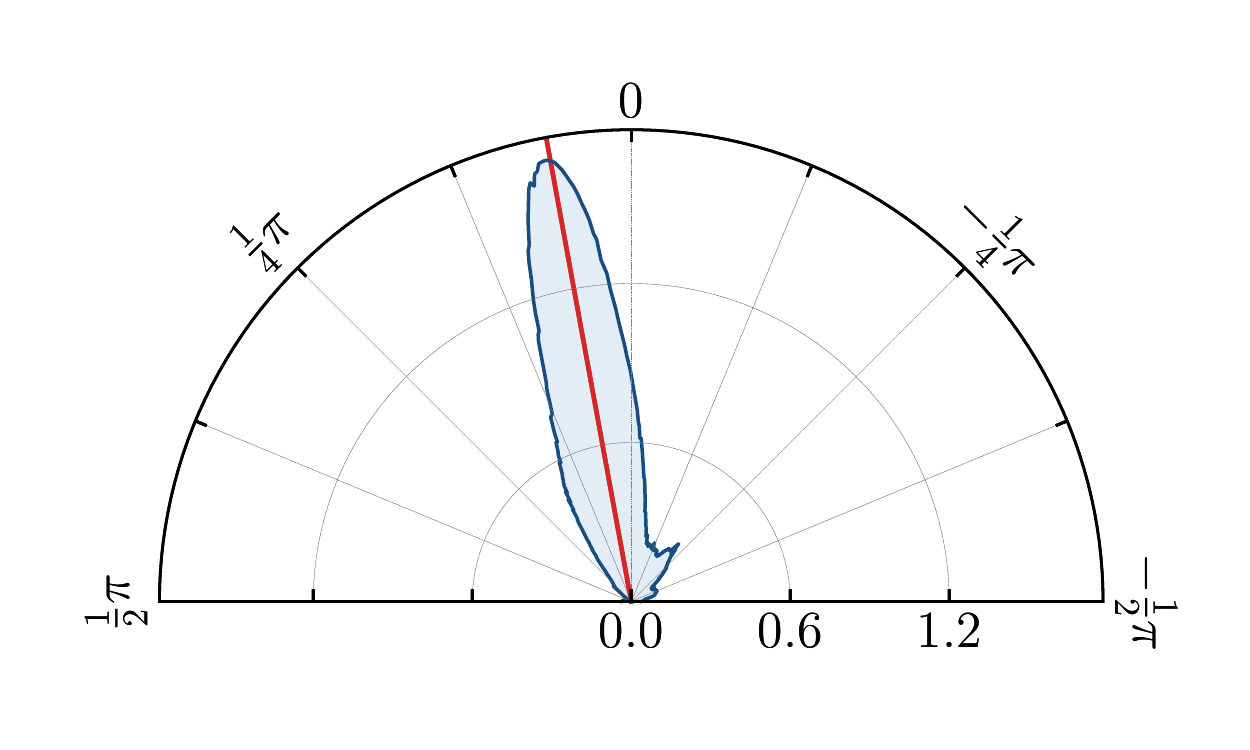}
\caption{Relative orientation of the H{\sc i} filaments in the 3C 196 field, identified with the RHT  ($Z=0.8$, $D_K=10'$ and $D_W=100'$). The red line marks a mean orientation.}
\label{fig:RHT_HI_Planck}
\end{figure}

The final distribution is presented in a half-polar plot given in Fig.~\ref{fig:RHT_HI_Planck}. The input parameters for the RHT analysis are $Z=0.8$, $D_K=10'$ , and $D_W=100'$.  Within the analyzed velocity range, the mean orientation of H{\sc i} filaments is $\left<\theta\right>=10^\circ$, and the spread of the distribution is $\delta\theta=\pm12^\circ$.  As demonstrated by \citet{clark14}, the results are robust to variation in the RHT parameters.

\section{Discussion and conclusions}\label{sec:dis}
Most of the straight depolarization canals detected in the 3C 196 field are inclined by $\sim 10^\circ$ with respect to the Galactic plane. They have the same orientation as the filament of the magneto-ionic medium, displacing the background synchrotron emission in Faraday depth.

The other distinct orientation of depolarization canals is $\sim -65^\circ$ with respect to the Galactic plane. They are associated with very weak emission observed in the upper part of the image presented in Fig.~\ref{fig:RHTimage}. These canals have similar orientation as the bar-like structure observed in the same region at higher radio frequencies \citep[WSRT observations at 350 MHz, presented in][see figure 10]{jelic15}. The bar-like structure is not detectable with LOFAR, since it is Faraday thick at LOFAR observing frequencies.

\begin{figure}[t]
    \centering \includegraphics[width=.75\linewidth]{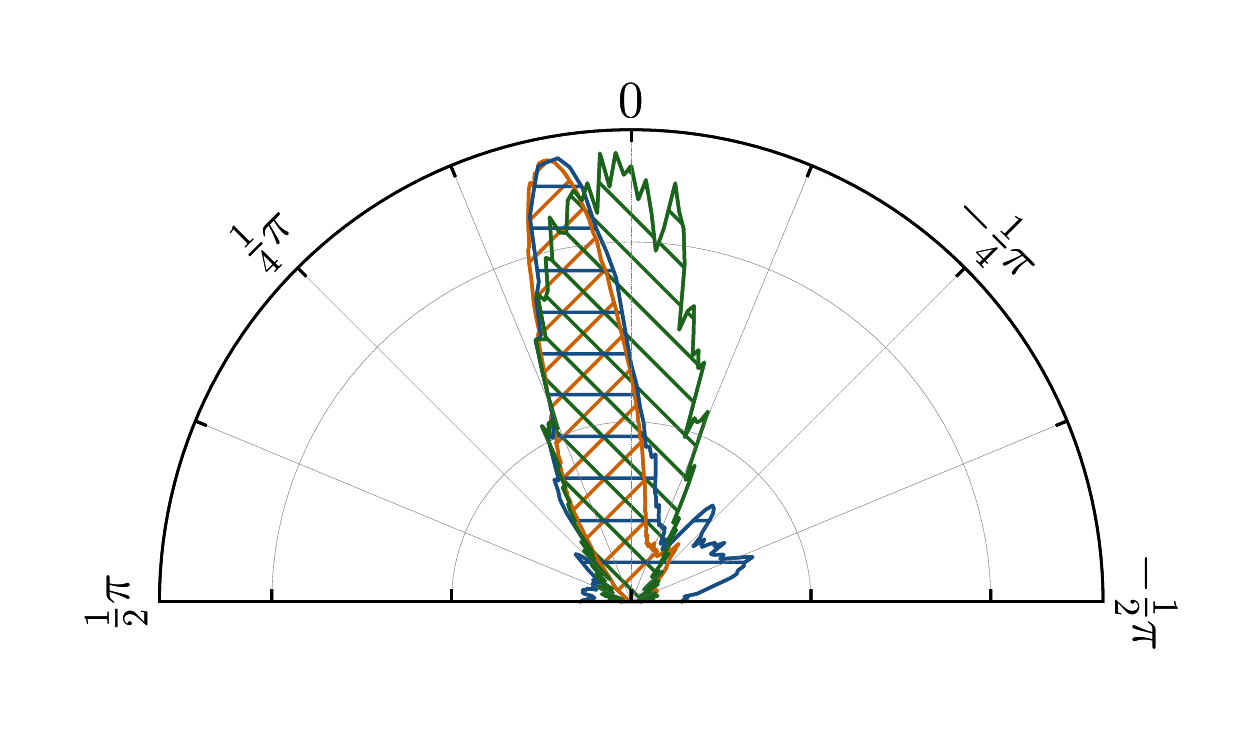}
    \caption{Comparison of the orientations of the straight depolarization canals (blue distribution, RHT input parameters: $Z=0.8$, $D_K=4'$ , and $D_W=30'$), the H{\sc i} filaments (orange distribution, RHT input parameters: $Z=0.8$, $D_K=10'$, and $D_W=100'$), and the plane-of-sky magnetic field component (green distribution) in the 3C 196 field. The scales are relative.}
    \label{fig:all_in_one}
\end{figure}

The orientation of the H{\sc i} filaments in the 3C 196 field matches that of the straight depolarization canals observed in the LOFAR polarimetric data. This is clearly visible in Fig.~\ref{fig:all_in_one}, which compares the two distributions given in relative scale. The H{\sc i} filaments have a dominant orientation of $\sim 10^\circ$ with respect to the Galactic plane at velocities from $-11.5~{\rm km~s^{-1}}$ to $+3.0~{\rm km~s^{-1}}$. Around velocities of $+15.0~{\rm km~s^{-1}}$, the H{\sc i} filaments located in the upper part of the field have a similar orientation as the $-65^\circ$ depolarization canals located in the same region (see Fig.~\ref{fig:RHT_HI_vel} in Appendix A, which shows the relative orientation of the H{\sc i} filaments as a function of  the line-of-sight velocity). 

The observed coherency of the H{\sc i} filament orientations over the wide rage of velocities indicates a very uniform and ordered magnetic field. On the other hand, the changing orientation of the neutral structures at higher velocities points to a tangled line-of-sight magnetic field \citep{clark18}.
 
In Figure~\ref{fig:all_in_one} we therfore also compare the orientations of the straight depolarization canals and the H{\sc i} filaments with the orientation of the plane-of-sky magnetic field component, whose distribution is presented in green. The three distributions are aligned. The magnetic field is on average almost parallel to the Galactic plane, with a mean orientation of $3^\circ$ and a spread of $\pm14^\circ$. 

This is consistent with the overall Galactic magnetic field direction. As discussed in \citet{jelic15}, the magnetic field component that follows the spiral arms of our Galaxy is almost perpendicular to the line of sight in the direction of 3C 196. Hence, it is expected that most of the structures detected in the LOFAR polarimetric data and in the H{\sc i} data follow the direction of the plane-of-sky magnetic field, in this case, the dominant orientation of the local magnetic field. 

However, we probe with the \emph{Planck} satellite the density-weighted and averaged magnetic field direction. Some local field at any given location along the line of sight can be obscured or averaged out in observations. This seems to be the case in the upper part of the 3C 196 field. An orientation of the local magnetic field associated with $-65^\circ$ structures is not observed in the \emph{Planck} data. 

This does not necessary imply a different dust-to-gas ratio of the related HI structures, localized around the velocities of $+15.0~{\rm km~s^{-1}}$. The observed magnetic field orientation in this region is simply dominated by a very coherent orientation of the magnetized H{\sc i} filaments, spanning a range of velocities from $-11.5~{\rm km~s^{-1}}$ to $+3.0~{\rm km~s^{-1}}$.

To summarize, the RHT is a robust method for characterizing the straight depolarization canals observed in radio-polarimetric data. It allow us to study their orientations largely independently of the RHT input parameters ($Z$, $D_K$ , and $D_W$). 

Based on the RHT analysis in the 3C~196 field, we find an alignment between three distinct tracers of the local ISM, of  (i) magneto-ionic structures and the associated depolarization canals observed in the LOFAR radio-polarimetric data, (ii) cold neutral filaments  observed in the H{\sc i}-EBHIS data, and (iii) the plane-of-sky magnetic field orientation probed by the \emph{Planck} 353 GHz polarization data. 

Our results support a picture that the ordered magnetic field plays a crucial role in confining different ISM phases. It also suggests that most of the observed straight depolarization canals in the 3C~196 field are indeed the result of a fortunate projection of the complicated three-dimensional distribution of the ISM.  In this particular part of the sky, its morphology is driven by the direction of a very ordered local magnetic field, which is in this case almost parallel to the Galactic plane. The magnetic field is also very coherent along the line of sight because the H{\sc i} filaments are aligned to each other over the wide range of velocities. 

These results are supported by the LOFAR follow-up observation (project code LC5\_008; Jeli\'c et al. in prep.) of a field located $\sim8^\circ$ below the 3C~196 field, toward the Galactic plane. This field shows multiple shell-like structures and straight depolarization canals parallel to the features in the 3C~196 field. The observed structures are also aligned with the \emph{Planck} plane-of-sky magnetic field orientation. This means that the 3C~196 field and this field might be part of the same very coherent large-scale magnetic field morphology, shaped by the common sources of the shock waves.

Moreover, because the three different ISM tracers are clearly connected, they can be used to constrain the relative distances to the observed structures in Faraday depth. For example, the Faraday depth components consistent with the $10^\circ$ HI filaments may be closer to us than the component consistent with the $-65^\circ$ HI filaments.

Finally, the RHT analysis also provides information on the position and the length of the identified linear structures. As we will show in a follow-up paper, this can be used to explore the possible association of some depolarization canals with the ionizing trails of stars. 

\begin{acknowledgements} 
We thank an anonymous referee for their constructive comments that improved the manuscript. This paper is based in part on data obtained with the International LOFAR Telescope (ILT) under project code LC0\_019. LOFAR \citep{haarlem13} is the Low Frequency Array designed and constructed by ASTRON. It has observing, data processing, and data storage facilities in several countries, that are owned by various parties (each with their own funding sources), and that are collectively operated by the ILT foundation under a joint scientific policy. The ILT resources have benefitted from the following recent major funding sources: CNRS-INSU, Observatoire de Paris and Universit\'e d\'Orl\'eans, France; BMBF, MIWF-NRW, MPG, Germany; Science Foundation Ireland (SFI), Department of Business, Enterprise and Innovation (DBEI), Ireland; NWO, The Netherlands; The Science and Technology Facilities Council, UK; Ministry of Science and Higher Education, Poland.
\end{acknowledgements} 

\bibliographystyle{aa}
\bibliography{reflist_polar}
\onecolumn
\begin{appendix}
\section{Relative orientation of the H{\sc i} filaments as a function of velocity.}
\begin{figure*}[h!]
    \centering \includegraphics[width=.24\linewidth]{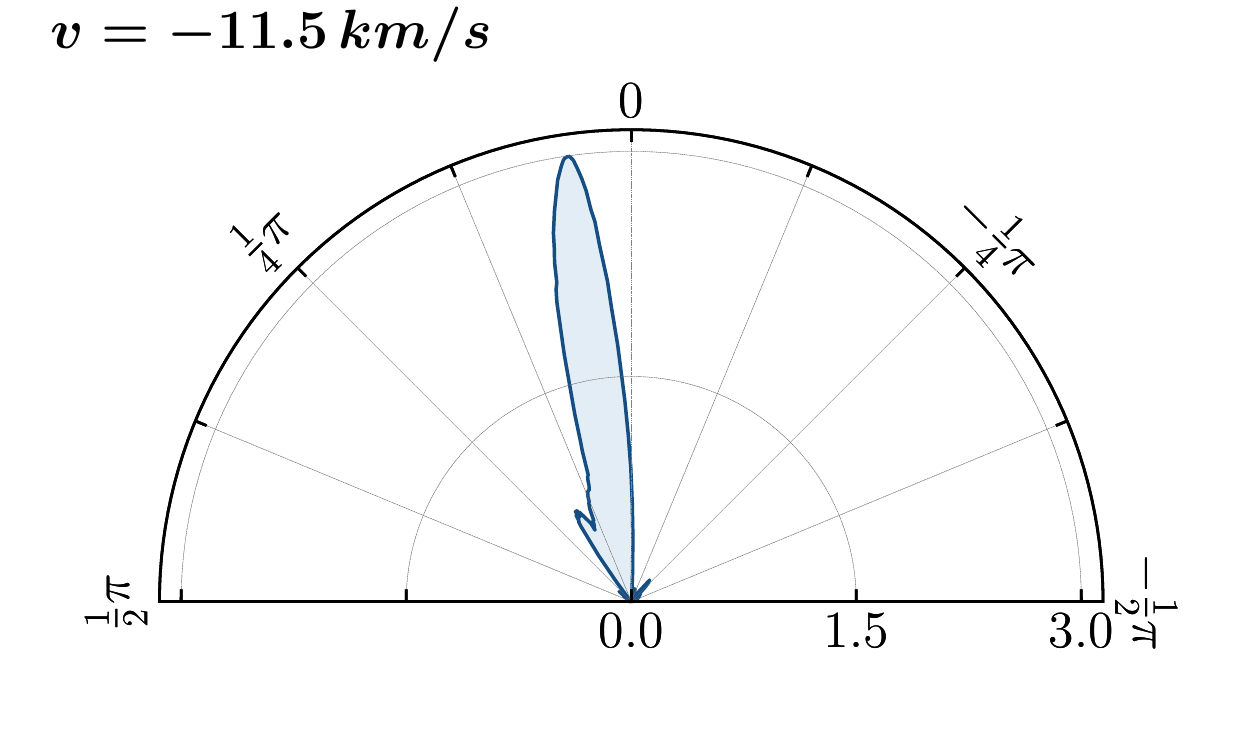}
    \centering \includegraphics[width=.24\linewidth]{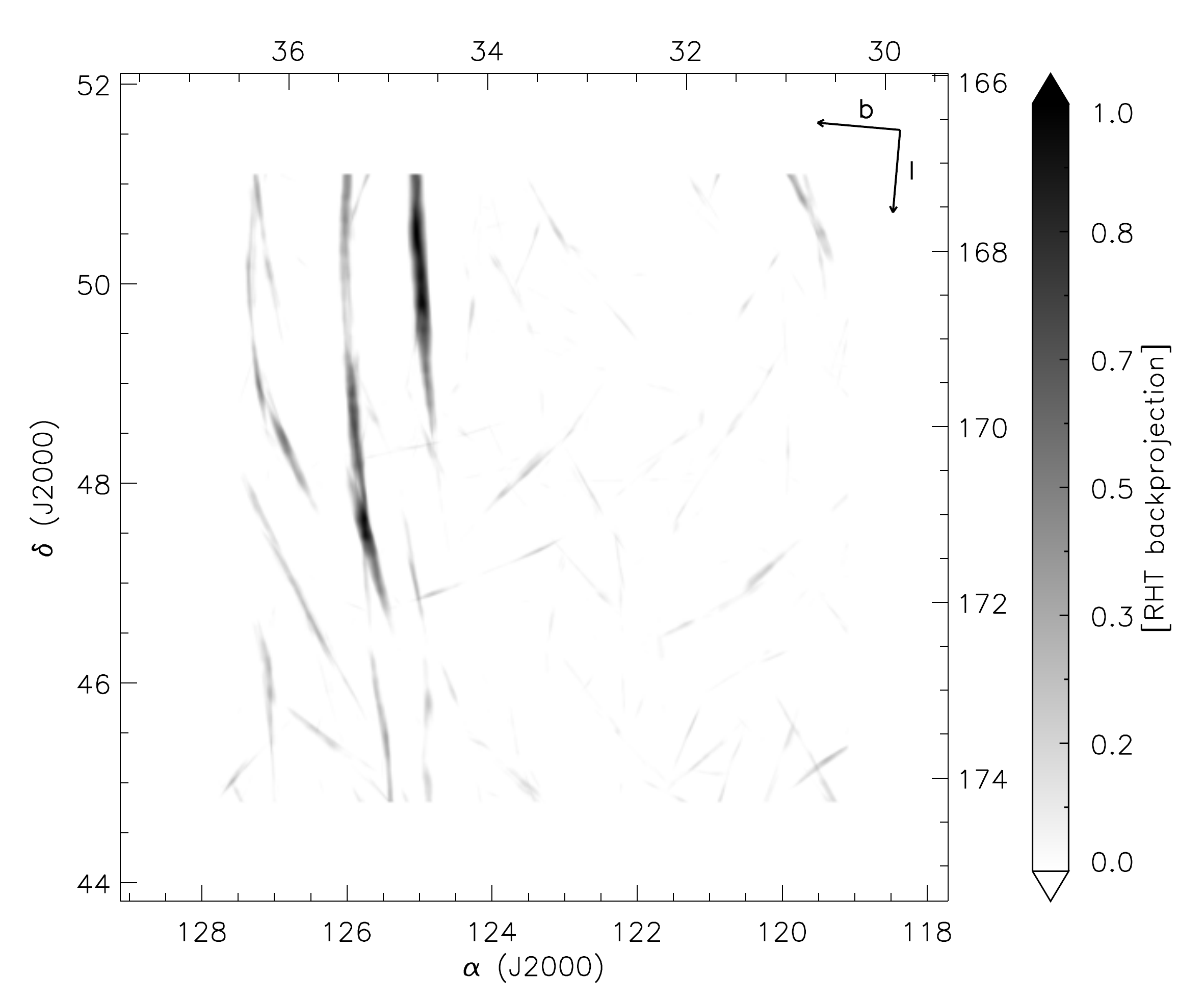}
    \centering \includegraphics[width=.24\linewidth]{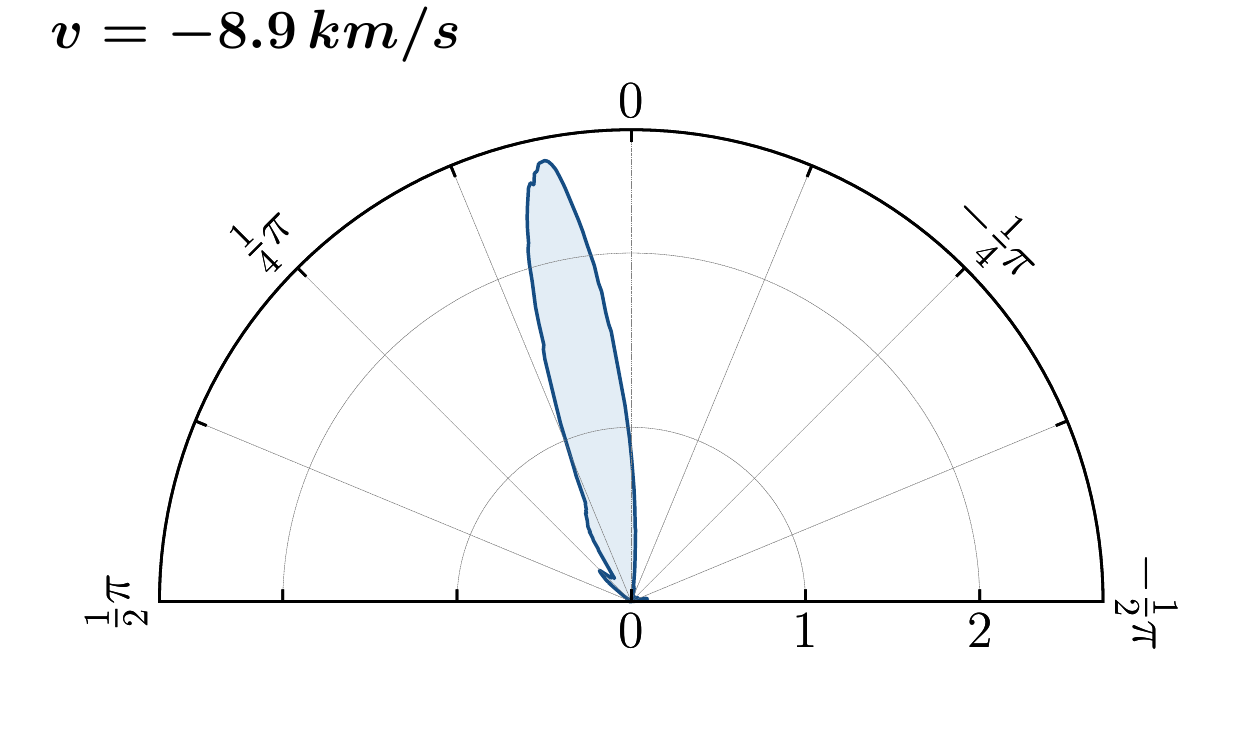}
    \centering \includegraphics[width=.24\linewidth]{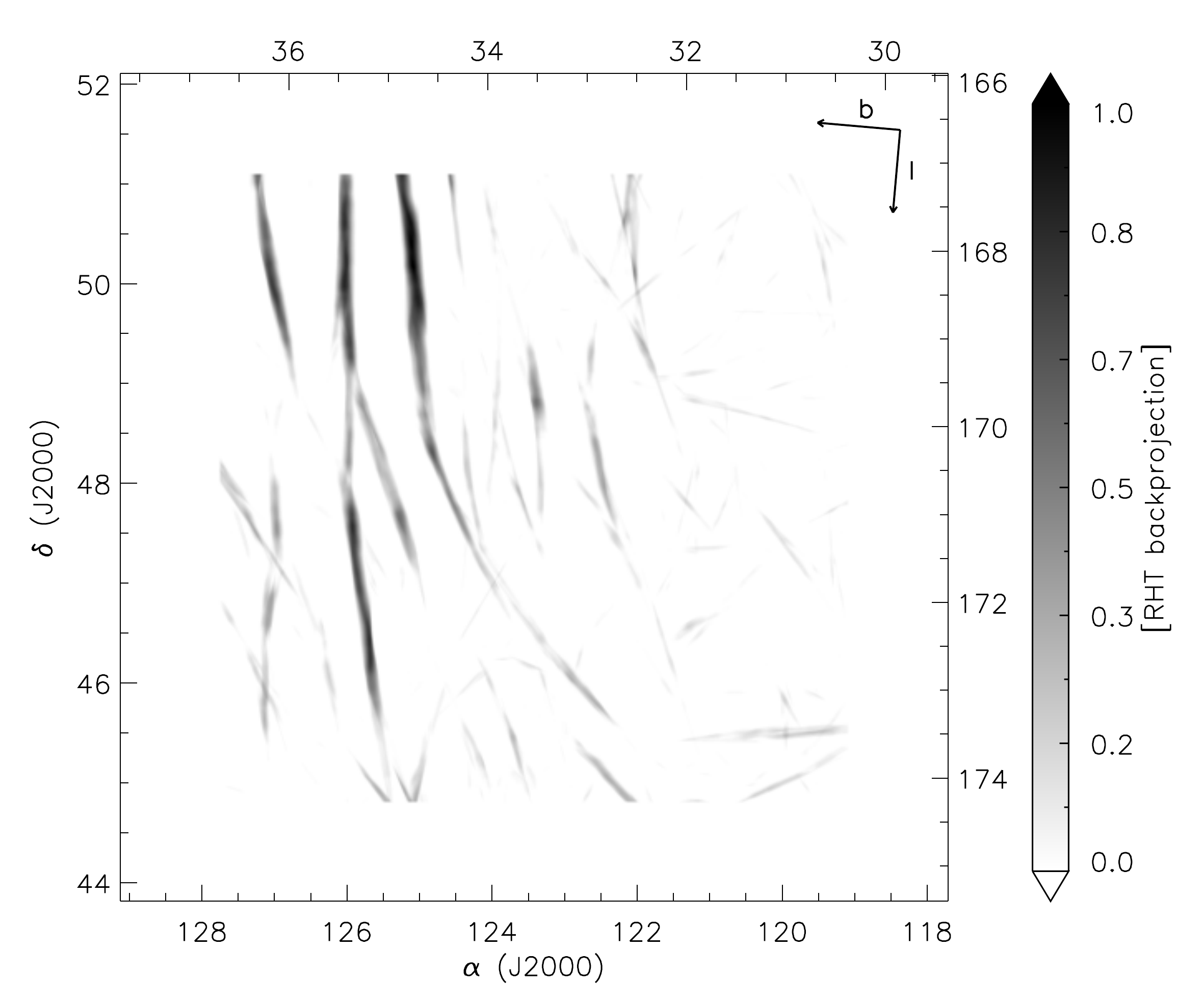}
    \centering \includegraphics[width=.24\linewidth]{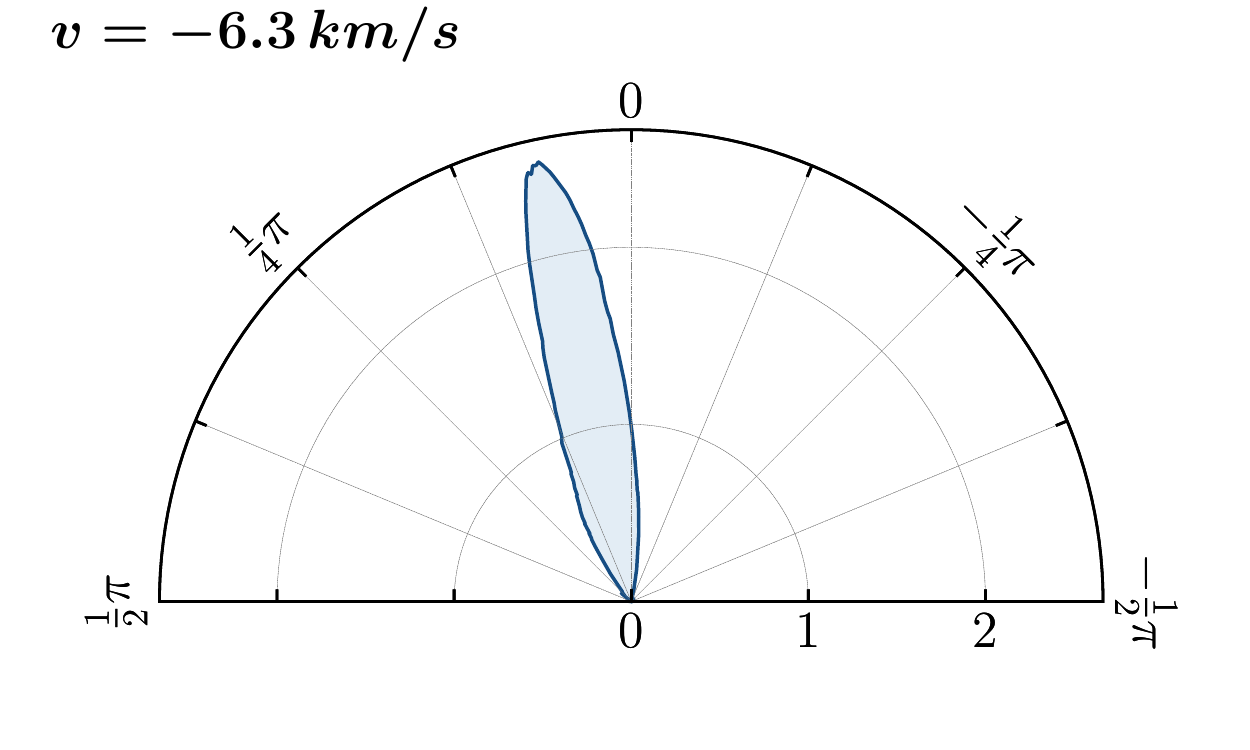}
    \centering \includegraphics[width=.24\linewidth]{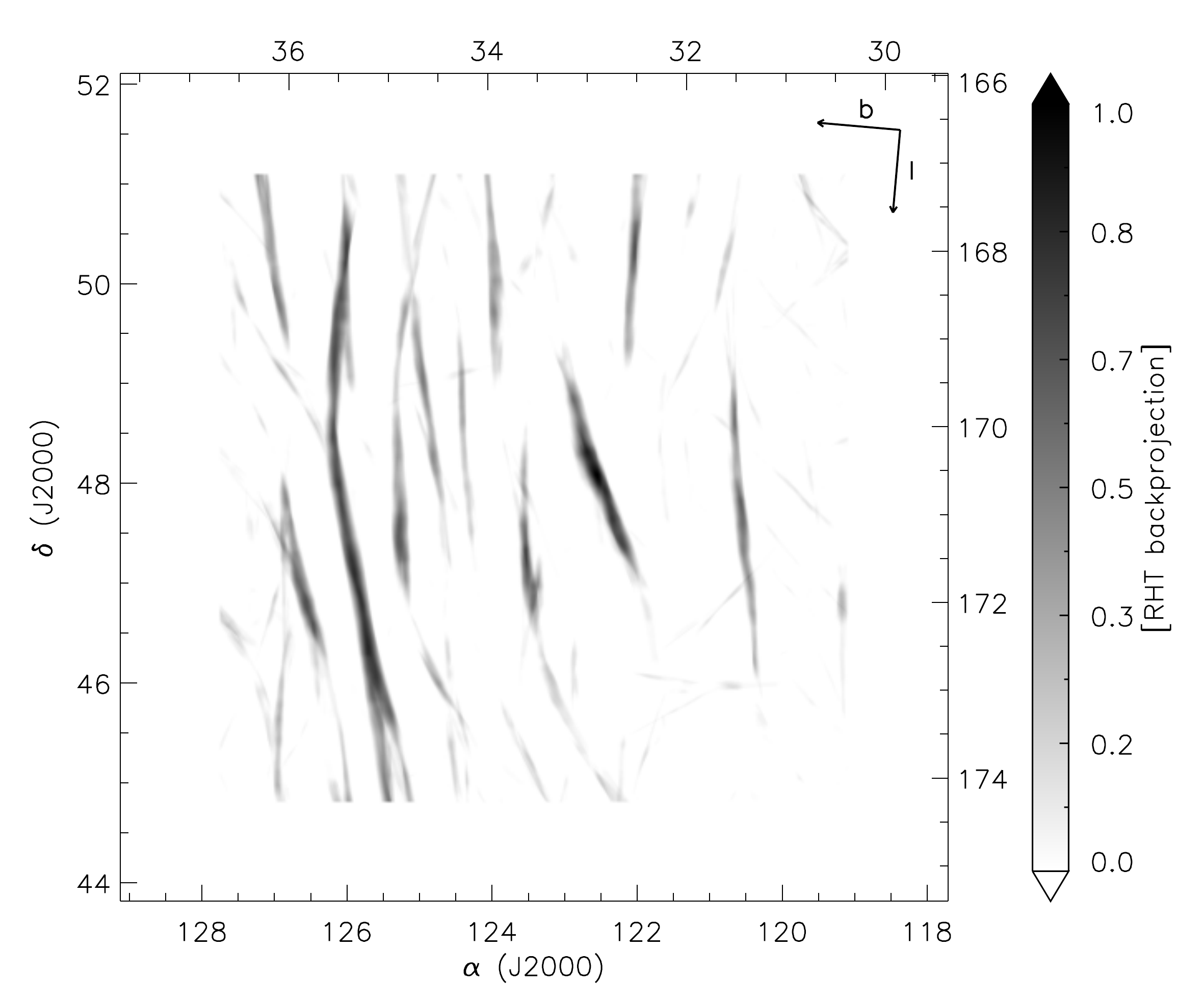}
    \centering \includegraphics[width=.24\linewidth]{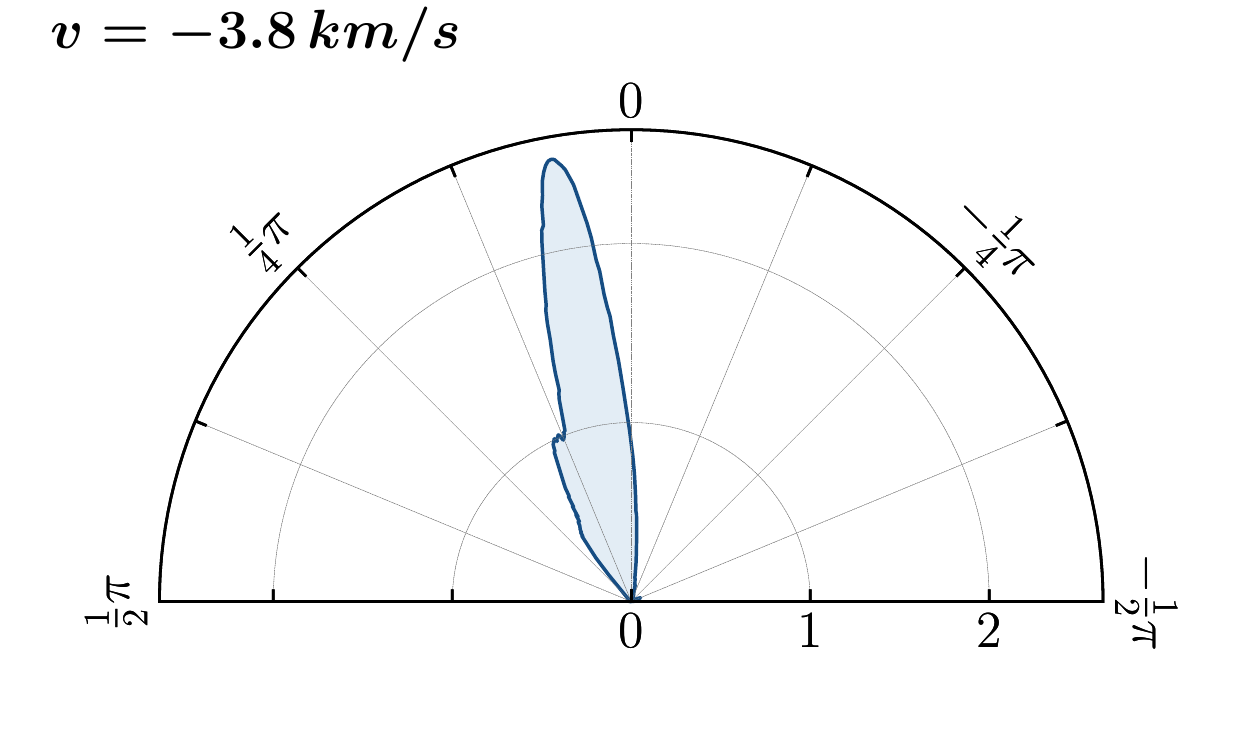}
    \centering \includegraphics[width=.24\linewidth]{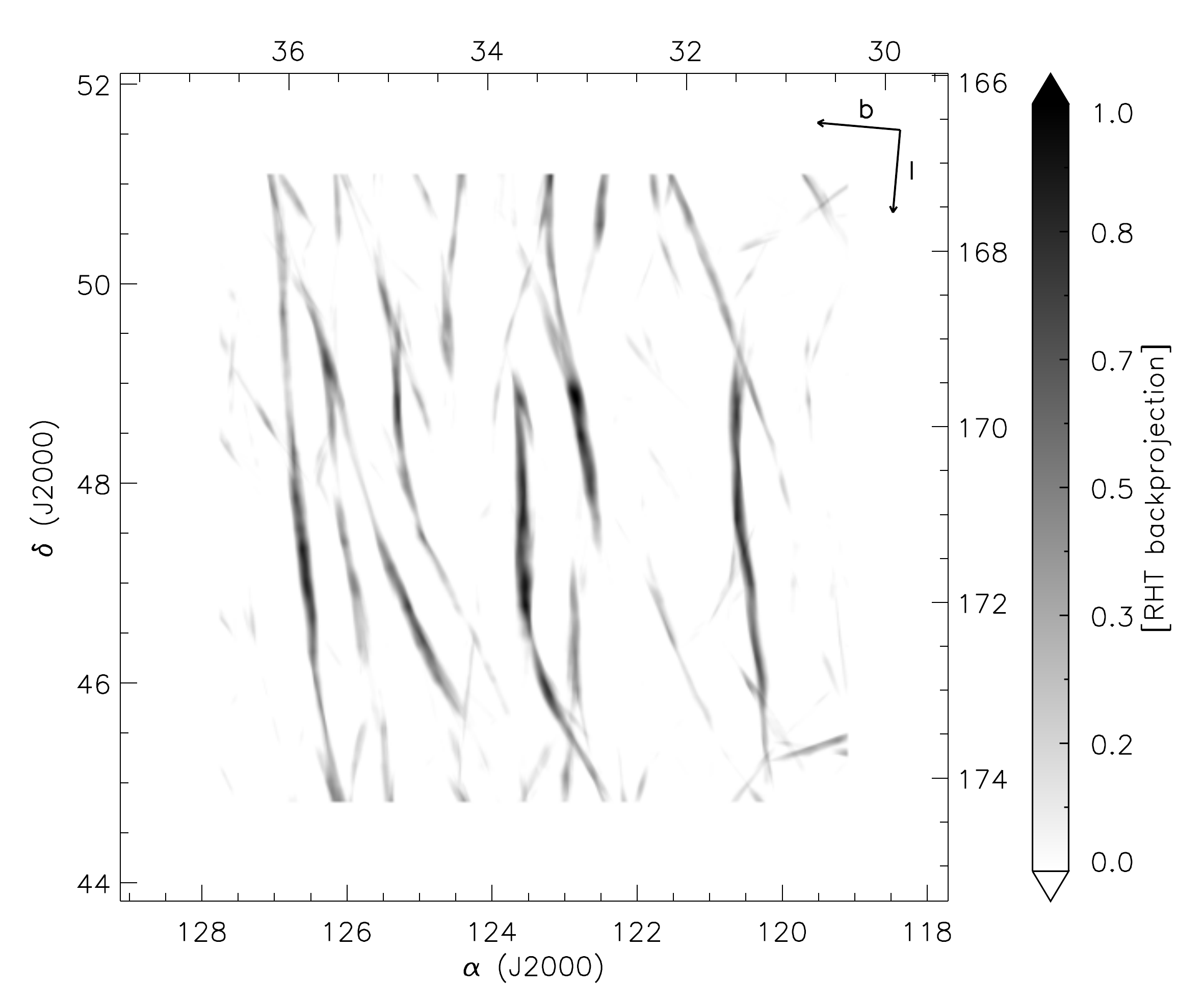}
    \centering \includegraphics[width=.24\linewidth]{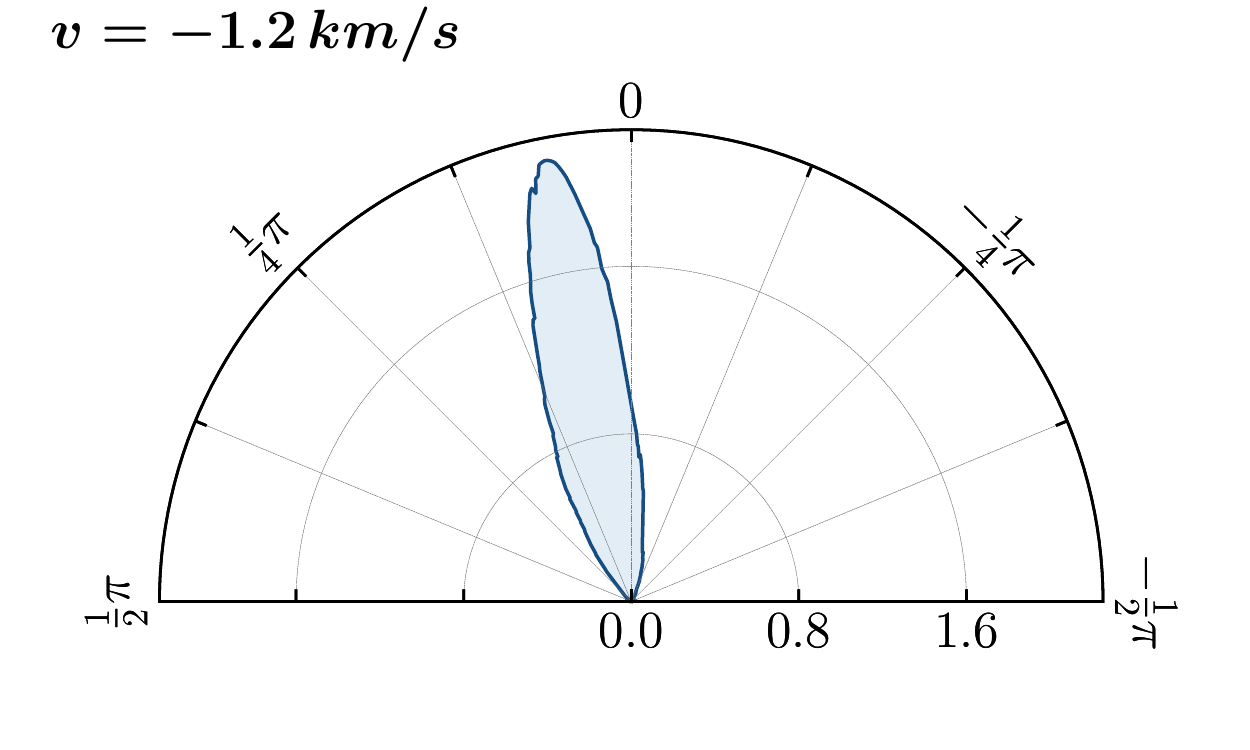}
    \centering \includegraphics[width=.24\linewidth]{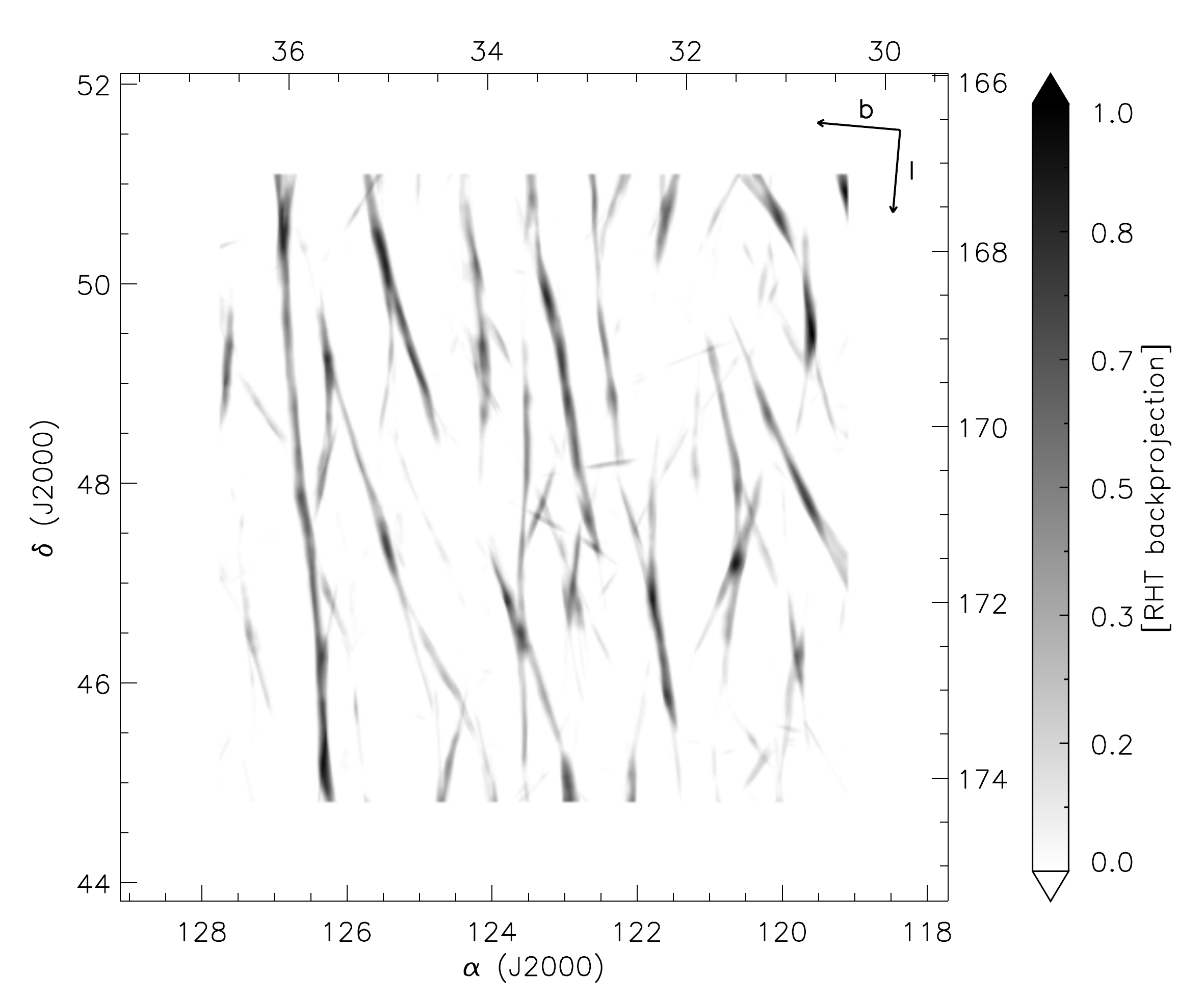}
    \centering \includegraphics[width=.24\linewidth]{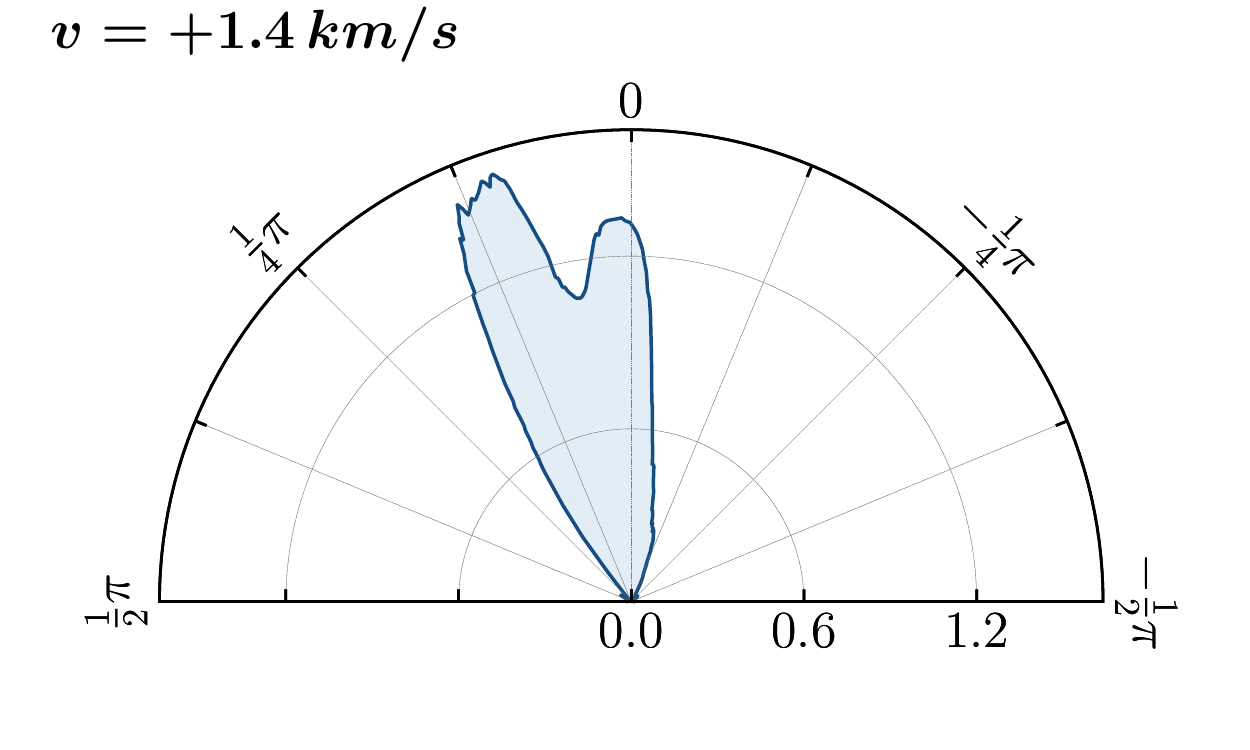}
    \centering \includegraphics[width=.24\linewidth]{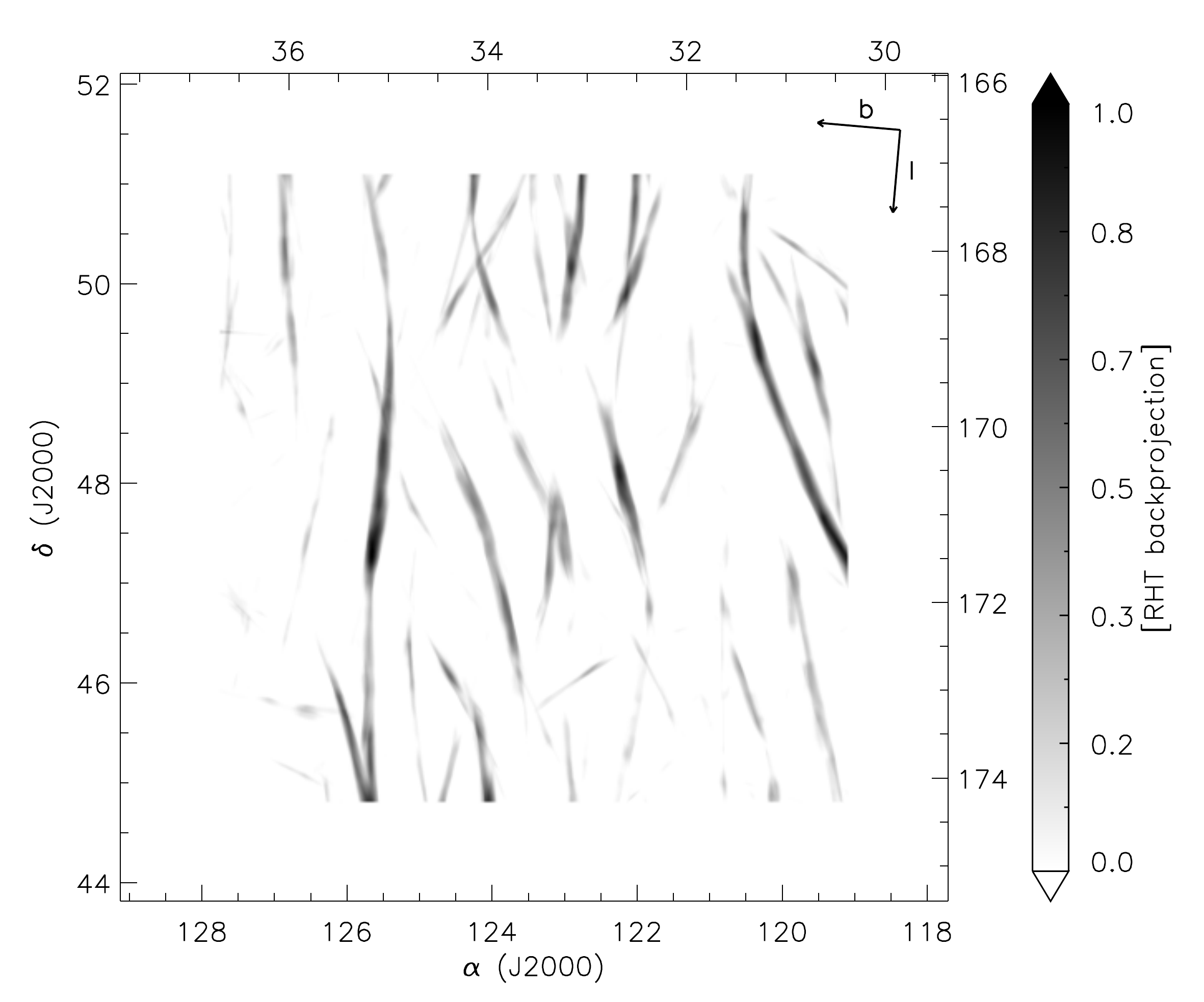}
    \centering \includegraphics[width=.24\linewidth]{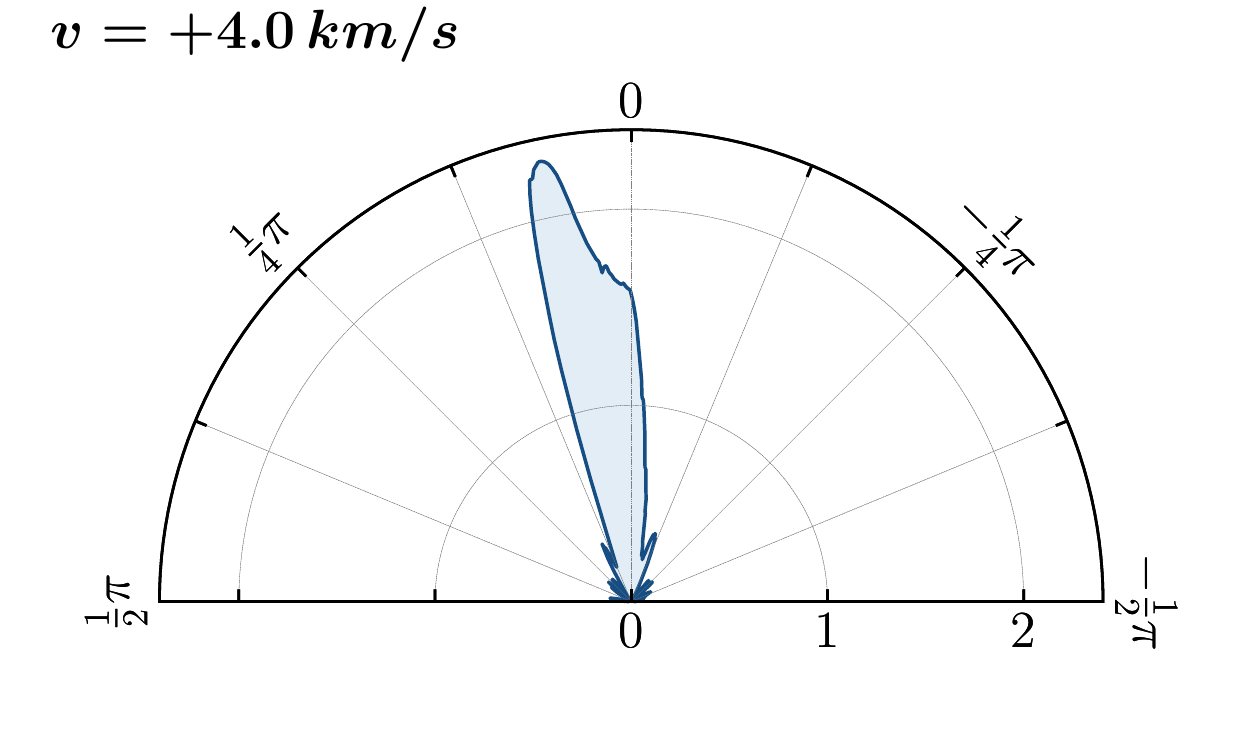}
    \centering \includegraphics[width=.24\linewidth]{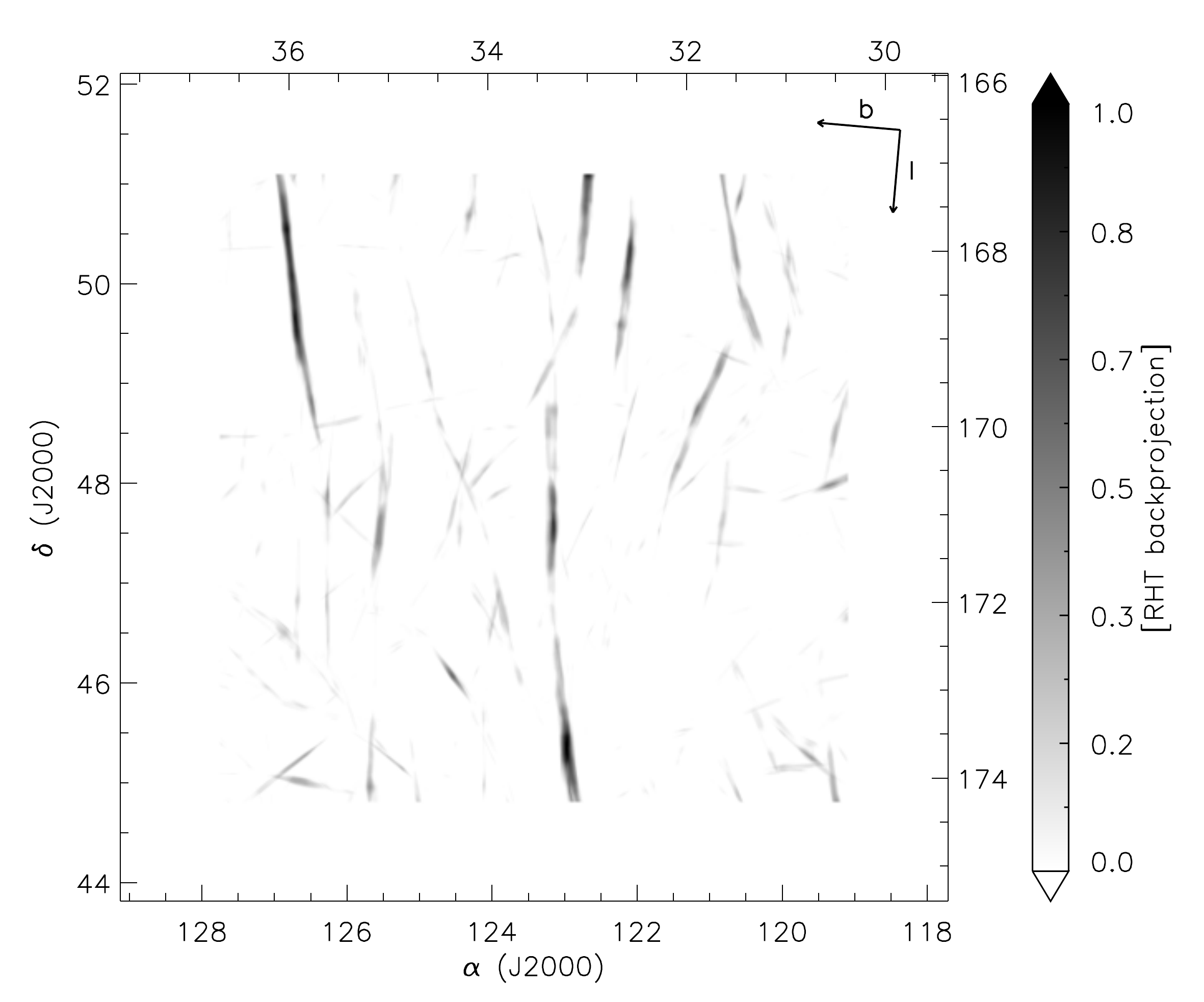}
    \centering \includegraphics[width=.24\linewidth]{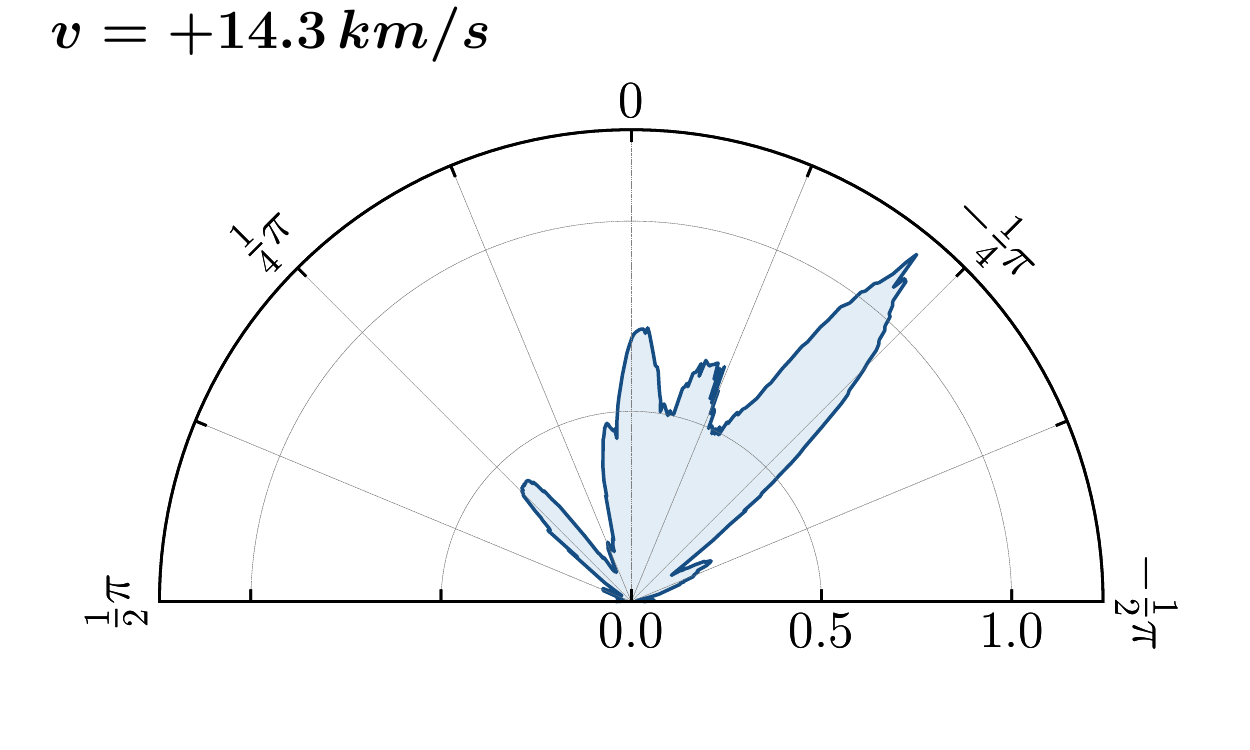}
    \centering \includegraphics[width=.24\linewidth]{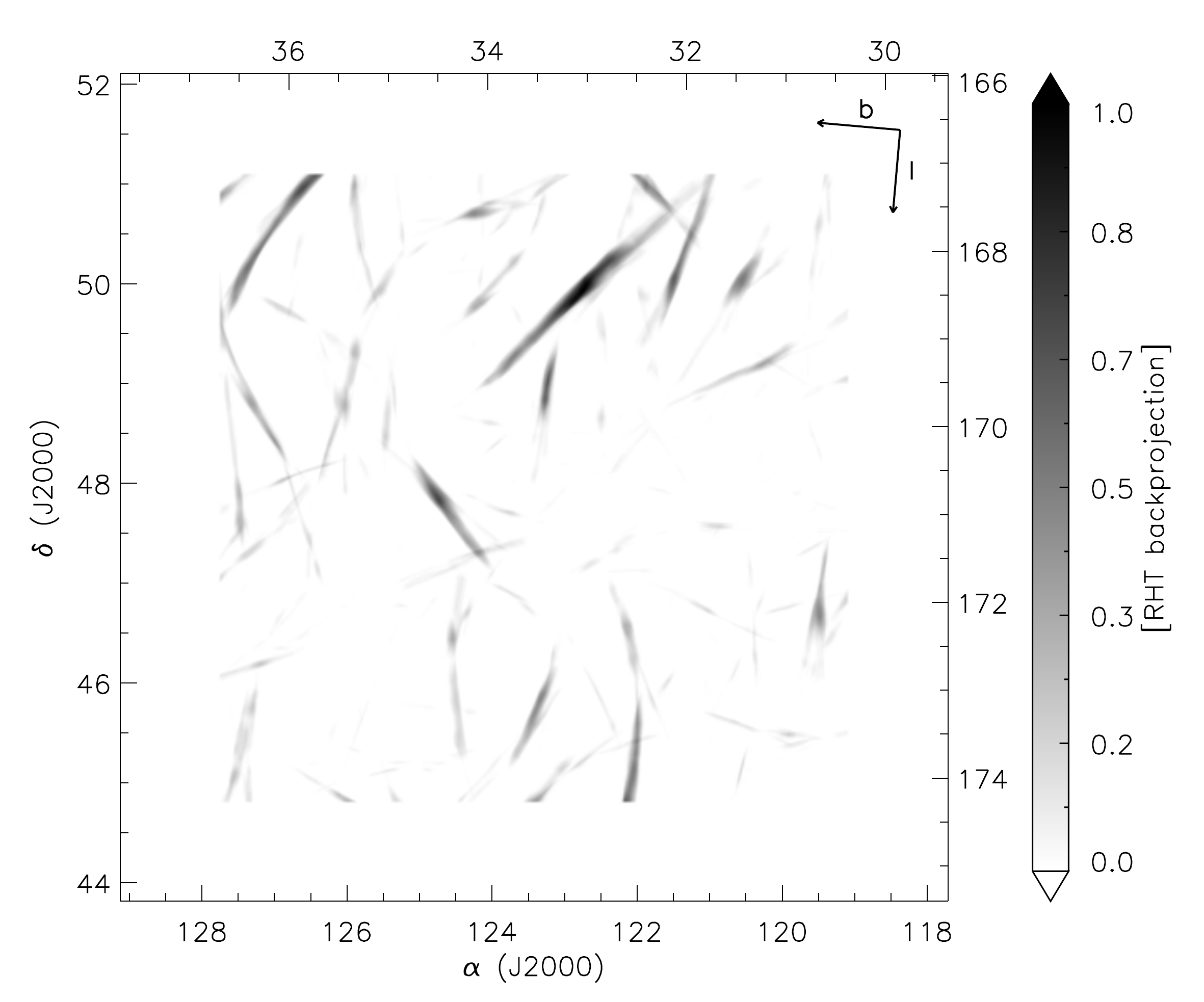}
    \centering \includegraphics[width=.24\linewidth]{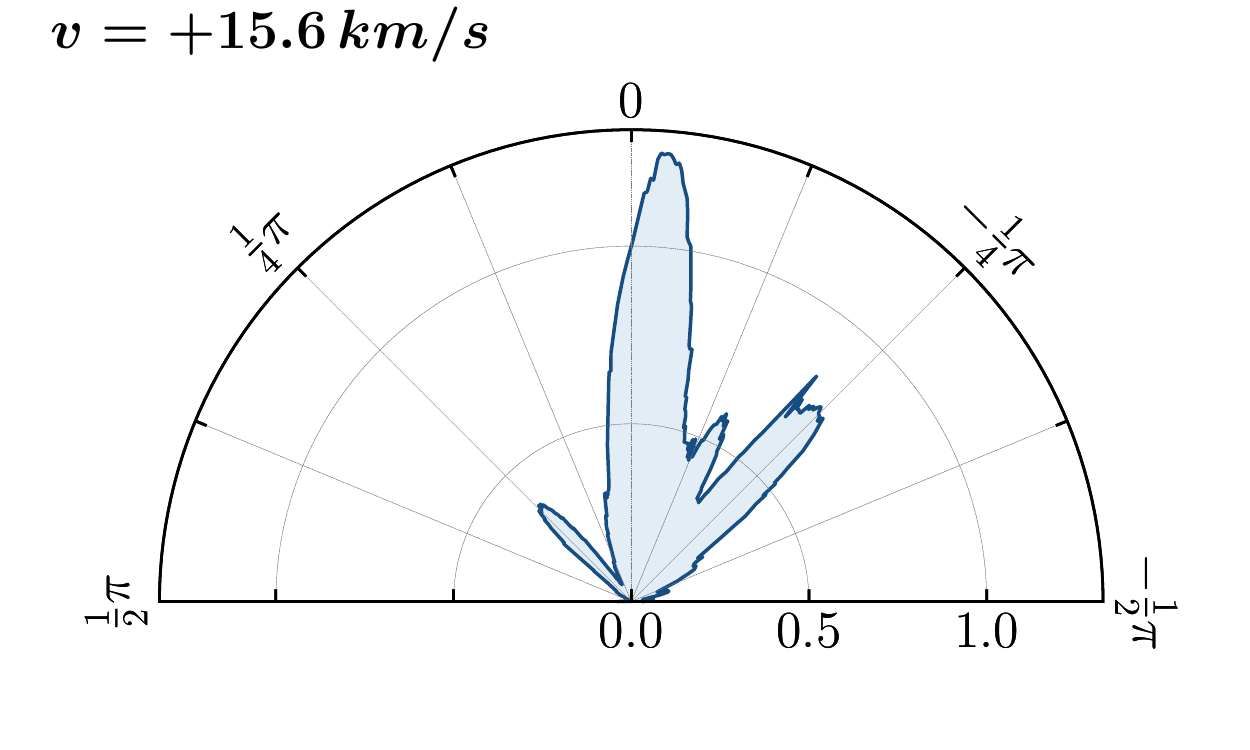}
    \centering \includegraphics[width=.24\linewidth]{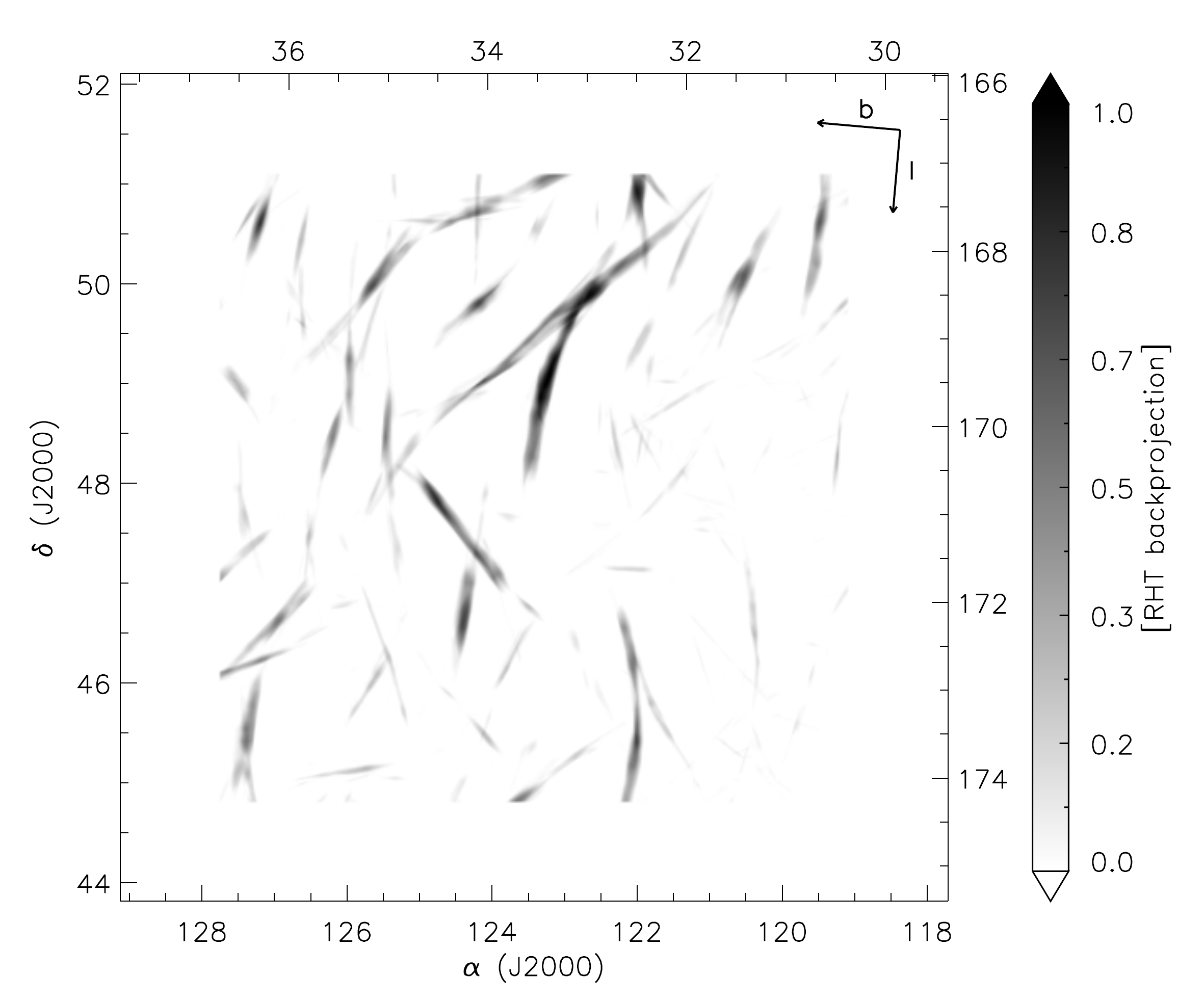}
    \centering \includegraphics[width=.24\linewidth]{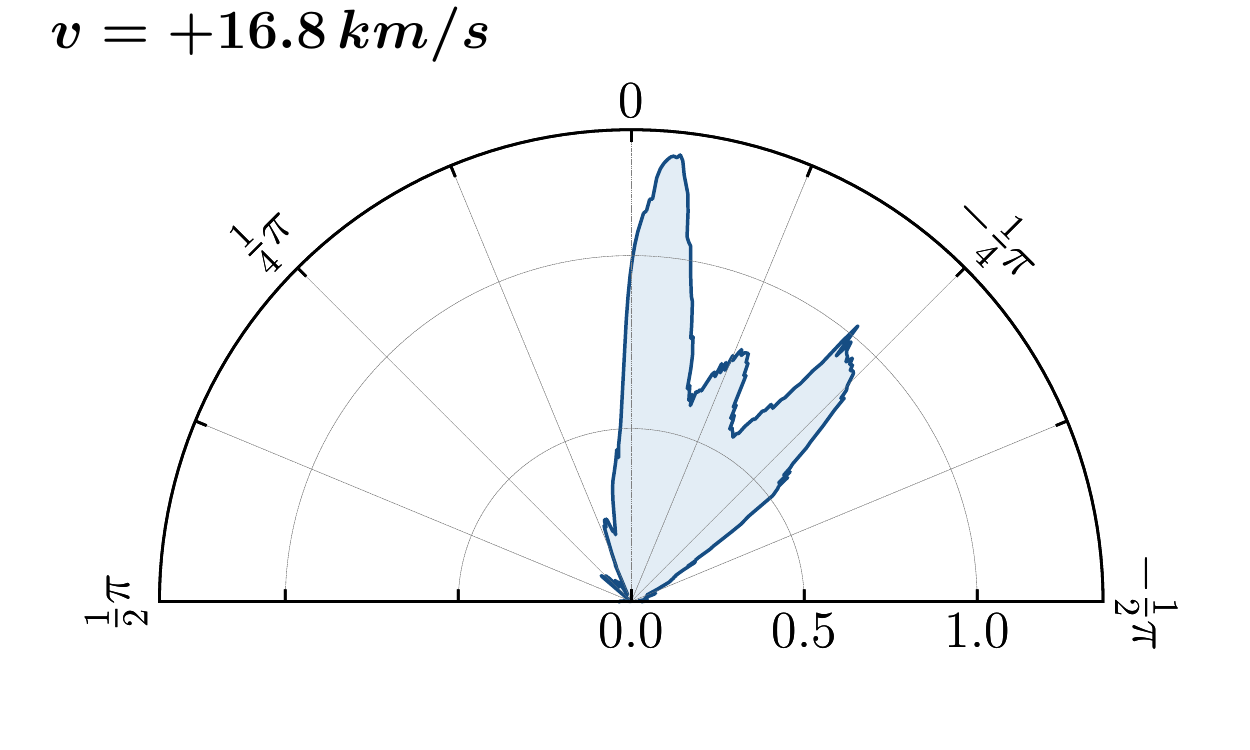}
    \centering \includegraphics[width=.24\linewidth]{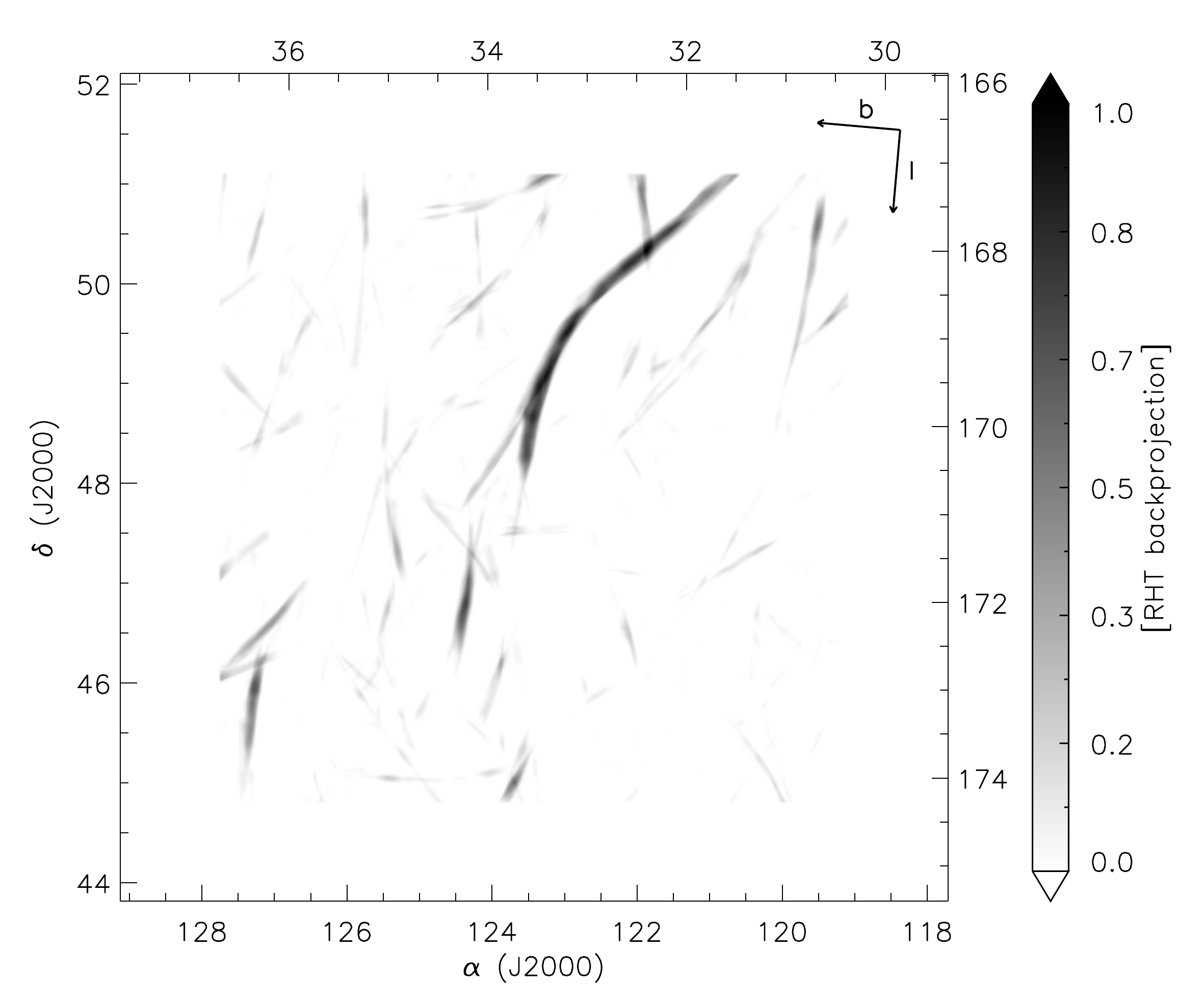}
    \caption{Relative orientation of the H{\sc i} filaments in the 3C 196 field for different velocity channels of the EBHIS data. 
    In addition to the half-polar plots, we show the RHT backprojections. The input parameters for the RHT are $Z=0.8$, $D_K=10'$ , and $D_W=100'$.}
    \label{fig:RHT_HI_vel}
\end{figure*}
\end{appendix}
\end{document}